\documentclass[11pt]{article}

\usepackage{amssymb} 
\usepackage{amsfonts} 
\usepackage{amsmath} 
\usepackage{latexsym} 
\usepackage{amsthm} 
\usepackage{hyperref}
\hypersetup{
     colorlinks   = true,
     citecolor    = blue,
pdfborderstyle={/S/U/W 1}
}
\usepackage{fullpage}
\usepackage{graphicx,graphics}
\usepackage{epsfig,color}
\usepackage{array,multirow}
\usepackage{colortbl,color}
\usepackage{floatflt}
\usepackage{shadethm}
\usepackage{mdframed}
\usepackage[usenames,dvipsnames]{xcolor}
\usepackage[T1]{fontenc}
\usepackage{fourier}
\usepackage{todonotes}

\usepackage{algorithm}
\usepackage[noend]{algpseudocode}

\usepackage{longtable}

\usepackage{pgf}

\usepackage{bm}
\usepackage{tikz}

\usepackage{pstricks,pst-node,pst-tree,pstricks-add}

\theoremstyle{plain}
\newtheorem{theorem}{Theorem}[section]
\newtheorem{definition}[theorem]{Definition}

\newtheorem{lemma}[theorem]{Lemma}

\newtheorem{corollary}[theorem]{Corollary}

\definecolor{orange}{rgb}{1,0.5,0}

\colorlet{LtOrange}{orange!50}

\newmdtheoremenv[outerlinewidth=1,leftmargin=40,%
rightmargin=40,backgroundcolor=LtOrange,%
outerlinecolor=LtOrange, innertopmargin=0pt,%
splittopskip=\topskip,skipbelow=\baselineskip,%
skipabove=\baselineskip]{ques}%
{Question}[section]

\colorlet{LtMagenta}{magenta!10}

\newmdtheoremenv[outerlinewidth=5,leftmargin=40,%
rightmargin=40,backgroundcolor=LtMagenta,%
outerlinecolor=blue, innertopmargin=0pt,%
splittopskip=\topskip,skipbelow=\baselineskip,%
skipabove=\baselineskip]{oques}%
{Open Question}[section]

\colorlet{LtMagenta}{blue!10}

\newmdtheoremenv[outerlinewidth=0,leftmargin=40,%
rightmargin=40,backgroundcolor=LtMagenta,%
outerlinecolor=blue, innertopmargin=0pt,%
splittopskip=\topskip,skipbelow=\baselineskip,%
skipabove=\baselineskip]{assum}%
{Assumption}[section]

\colorlet{LtYellow}{yellow!10}

\global\mdfdefinestyle{Problem}{%
linecolor=LtYellow, leftmargin=1cm, rightmargin=1cm, backgroundcolor=LtYellow}

\newcommand{\prob}[2]{
\vspace*{2mm}
\begin{mdframed}[style=Problem]
\begin{itemize}
\item \textbf{Input}: {#1}
\item \textbf{Output}: {#2}
\end{itemize}
\end{mdframed}
\vspace*{2mm}
}

\colorlet{LtGreen}{green!10}

\global\mdfdefinestyle{Algo}{%
linecolor=LtGreen, leftmargin=1cm, rightmargin=1cm, backgroundcolor=LtGreen}

\algrenewcommand\algorithmicrequire{\textsc{Input:}}
\algrenewcommand\algorithmicensure{\textsc{Output:}}
\algrenewcommand\algorithmicwhile{\textsc{while}}
\algrenewcommand\algorithmicfor{\textsc{for}}
\algrenewcommand\algorithmicdo{\textsc{do}}
\algrenewcommand\algorithmicreturn{\textsc{return}}
\algrenewcommand\algorithmicif{\textsc{if}}
\algrenewcommand\algorithmicthen{\textsc{then}}
\algrenewcommand\algorithmicelse{\textsc{else}}

\colorlet{LtRed}{red!10}

\global\mdfdefinestyle{Note}{%
linecolor=LtRed, leftmargin=1cm, rightmargin=1cm, backgroundcolor=LtRed}

\newcommand{\note}[1]{
\begin{mdframed}[style=Note]
{#1}
\end{mdframed}
}

\newcommand{\ignore}[1]{}

\renewcommand{\P}{\mathsf{P}}
\newcommand{\NP}{\mathsf{NP}}

\newcommand{\eps}{\varepsilon}
\renewcommand{\epsilon}{\varepsilon}

\newcommand{\vM}{\mathbf{M}}
\newcommand{\ve}{\mathbf{e}}
\newcommand{\vE}{\mathbf{E}}

\newcommand{\vF}{\mathbf{F}}

\newcommand{\vC}{\mathbf{C}}

\newcommand{\vB}{\mathbf{B}}
\newcommand{\va}{\mathbf{a}}
\newcommand{\vA}{\mathbf{A}}
\newcommand{\vd}{\mathbf{d}}
\newcommand{\vD}{\mathbf{D}}
\newcommand{\vH}{\mathbf{H}}
\newcommand{\vI}{\mathbf{I}}

\newcommand{\vy}{\mathbf{y}}

\newcommand{\vx}{\mathbf{x}}
\newcommand{\vhx}{\hat{\mathbf{x}}}
\newcommand{\vX}{\mathbf{X}}
\newcommand{\vz}{\mathbf{z}}
\newcommand{\vZ}{\mathbf{Z}}
\newcommand{\vv}{\mathbf{v}}
\newcommand{\vV}{\mathbf{V}}

\newcommand{\vW}{\mathbf{W}}
\newcommand{\vu}{\mathbf{u}}

\newcommand{\vP}{\mathbf{P}}
\newcommand{\vL}{\mathbf{L}}
\newcommand{\vR}{\mathbf{R}}

\newcommand{\vs}{\mathbf{s}}
\newcommand{\vS}{\mathbf{S}}
\newcommand{\vt}{\mathbf{t}}
\newcommand{\vT}{\mathbf{T}}

\newcommand{\vtheta}{\bm{\theta}}

\newcommand{\vzero}{\vec{0}}

\newcommand{\F}{\mathbb{F}}

\newcommand{\R}{\mathbb{R}}
\newcommand{\C}{\mathbb{C}}

\newcommand{\cC}{\mathcal{C}}

   \newcommand{\be}{\begin{equation}}
   \newcommand{\ee}{\end{equation}}

\newcolumntype{K}{>{\columncolor[rgb]{1,.75,.793}}m}
\newcolumntype{G}{>{\columncolor[rgb]{0,.5,0}}m}
\newcolumntype{R}{>{\columncolor[rgb]{0.5,0,0}}m}
\newcolumntype{Z}{>{\columncolor[rgb]{0,1,1}}m}
\newcolumntype{Y}{>{\columncolor{orange}}m}
\newcolumntype{A}{>{\centering\arraybackslash}K{40pt}}
\newcolumntype{B}{>{\centering\arraybackslash}m{40pt}}
\newcolumntype{C}{>{\centering\arraybackslash}R{40pt}}
\newcolumntype{D}{>{\centering\arraybackslash}G{40pt}}
\newcolumntype{E}{>{\centering\arraybackslash}Z{40pt}}
\newcolumntype{F}{>{\centering\arraybackslash}Y{40pt}}

\newcommand{\ip}[2]{\left\langle {#1},{#2}\right\rangle}

\newcommand{\paren}[1]{\left({#1}\right)}

\newcommand{\relu}{\textsf{ReLu}}

\newcommand{\tO}[1]{\widetilde{O}\left({#1}\right)}

\newcommand{\deriv}[2]{\nabla_{{#1}}\left({#2}\right)}

\newcommand{\norm}[2]{\left\| {#2}\right\|_{{#1}}}

\newcommand{\diag}{\text{diag}}

\newcommand{\propone}{\textsc{Expressivity property}}
\newcommand{\proptwo}{\textsc{Efficient MVM property}}
\newcommand{\propthree}{\textsc{Efficient gradient property}}

\usepackage{amsmath,amsfonts,bm}

\def\eqref#1{equation~\ref{#1}}

\def\ceil#1{\lceil #1 \rceil}

\def\1{\bm{1}}

\def\eps{{\epsilon}}

\def\vzero{{\bm{0}}}

\def\vtheta{{\bm{\theta}}}
\def\va{{\bm{a}}}

\def\vd{{\bm{d}}}
\def\ve{{\bm{e}}}

\def\vs{{\bm{s}}}
\def\vt{{\bm{t}}}
\def\vu{{\bm{u}}}
\def\vv{{\bm{v}}}

\def\vx{{\bm{x}}}
\def\vy{{\bm{y}}}
\def\vz{{\bm{z}}}

\DeclareMathAlphabet{\mathsfit}{\encodingdefault}{\sfdefault}{m}{sl}
\SetMathAlphabet{\mathsfit}{bold}{\encodingdefault}{\sfdefault}{bx}{n}

\newcommand{\B}{\mathcal{B}}
\newcommand{\BBS}{\B\B^*}

\usepackage{figures}

\title{\textbf{Arithmetic Circuits, Structured Matrices and (not so) Deep Learning}}
\author{\textsc{Atri Rudra}}
\date{Department of Computer Science and Engineering\\
University at Buffalo\\
\texttt{atri@buffalo.edu}}

\begin{document}

\maketitle




\begin{abstract}
This survey presents a necessarily incomplete (and biased) overview of results at the intersection of arithmetic circuit complexity, structured matrices and deep learning. Recently there has been some research activity in  replacing unstructured weight matrices in neural networks by structured ones (with the aim of reducing the size of the corresponding deep learning models). Most of this work has been experimental and in this survey, we formalize the research question and show how a recent work that combines arithmetic circuit complexity, structured matrices and deep learning essentially answers this question.

This survey is targeted at complexity theorists who might enjoy reading about how tools developed in arithmetic circuit complexity helped design (to the best of our knowledge) a new family of structured matrices, which in turn seem well-suited for applications in deep learning. However, we hope that folks primarily interested in deep learning would also appreciate the connections to complexity theory.
\end{abstract}

\newpage

\begin{center}
\textsc{In Memory of Alan Selman}
\begin{quote}
{\em
Alan Selman was my colleague at University at Buffalo (UB) from 2007 (when I joined UB) until 2014 (when Alan retired from UB). I still remember being taken directly from the airport to Alan’s favorite restaurant, Trattoria Aroma, during my interview at Buffalo. I was a bit intimidated by Alan during the dinner but we bonded over the fact that we were both married to epidemiologists. After I joined Buffalo, Alan’s sage advice helped me throughout my tenure process. Alan was a giant in the department and having him in my corner did not hurt. 

More germane to this survey, Alan always turned up for UB theory meetings and I greatly enjoyed presenting stuff I was working on to Alan during some of these meetings. After Alan retired in 2014, I started working on some problems at the intersection of arithmetic circuit complexity, structured matrices and deep learning that I think Alan would be enjoyed hearing about. Since Alan passed in early 2021, this survey is my way of presenting the material to Alan in his memory. \hfill{-- \textsc{Atri Rudra}}
}
\end{quote}
\end{center}

\section{Introduction}

This survey shows how concepts in arithmetic circuit complexity and structured matrices can be used to solve a (theoretical) problem motivated by practical applications in machine learning (especially deep learning). 
Since each of the areas of arithmetic (circuit) complexity, structured matrices and deep learning have been explored in great depth and this survey clearly cannot do any justice to all the great work in each of the these areas, we will spend most of the introduction clarifying what this survey is {\em not} about.

\paragraph{Algebraic circuit complexity} or more generally algebraic complexity theory~\cite{burgisser2013algebraic} studies the power of algebraic algorithms (as opposed to the Turing machine/RAM model). The arithmetic circuit model (or the straight-line programs) are one of the standard models of computation in algebraic complexity theory~\cite[Chapter 4]{burgisser2013algebraic}. 
In this survey we will ignore pretty much everything in this literature except for results on the arithmetic circuit complexity of the linear map i.e. functions of the form $\vx\mapsto \vW\vx$ (where $\vx$ is a vector over some field $\F$ and $\vW$ is a matrix over the same field)~\cite[Chapter 13]{burgisser2013algebraic}. We would like to stress that this survey will only scratch the surface of the literature on the algebraic circuit complexity of the linear map. Just to give a sense of the breadth of this seemingly `specialized' topic, we remark that the study of {\em matrix rigidity}~\cite{matrix-rig}, which has seen a lot of recent research activity~\cite{mat-rig-1,mat-rig-2,mat-rig-3,DL20,mat-rig-4}, is a part of this topic. We note that originally, the topic of matrix rigidity was proposed by Valiant~\cite{Val77} as a way to prove super-linear lower bounds, by constructing matrices that are {\em rigid}. However, our goal in this survey is to prove upper bounds-- i.e. we are interested in matrices for which the arithmetic circuit complexity is small. We note that some of the recent work, including  the work of Alman and Williams~\cite{mat-rig-1}, is along similar lines of showing that explicit matrices are {\em not} rigid (which at a very hand-wavy level is showing that for certain explicit linear maps there indeed exist `small' arithmetic of a {\em restricted kind} to compute the linear map\footnote{The notion of small here is to show circuits of size $o(n^2)$ but the bounds are still $\Omega\paren{n^{2-\eps}}$ for any fixed $\eps>0$, while in our case we are more interested in linear maps that have a near-linear sized {\em general} arithmetic circuits.}
-- see Section~\ref{sec:mat-rig} for more details).

\paragraph{Structured matrices} are (family) of matrices $\vW$ for which one can have a much smaller representation than the generic $n\times n$ (assuming $\vW$ is a square) matrix representation. Typically, these structured representations also imply that one can compute $\vW\vx$ for any vector $\vx$ in $o(n^2)$ time (recall that matrix-vector multiplication in the worst-case takes $O(n^2)$ time), and in many celebrated examples (e.g. FFT for the Discrete Fourier matrix~\cite{fft}), it takes near-linear time/operations over the underlying field. At the risk of over-simplifying things, structured matrices crop up in applications in two flavors. In the first flavor, the application essentially determines the (family) of structured matrices. In other words, we do not have any say in the choice of the structured matrix and the goal is design efficient matrix vector multiplication algorithm (or algorithm for some other problems) involving the matrix. We mention two examples. The first example is the family of  {\em orthogonal polynomial transforms} (which include among others the Discrete Cosine Transform)~\cite{szego} that appear in many signal processing applications as well as many basic mathematical studies including approximation theory. The second example is the family of  {\em low displacement rank} matrices~\cite{kailath1979displacement}, which have applications in signal processing and numerical linear algebra~\cite{k-mat-survey}. We will not cover this flavor of structured matrices in the survey though low displacement rank matrices will make an appearance in Section~\ref{sec:ldr}.

The second flavor of structured matrices (which we will focus on in this survey), is where there is some matrix $\vM$ in the `wild' and we want to approximate $\vM$ by a more structured matrix $\vW$ to e.g. save on storing the matrix (and/or have more efficient operations on the matrix: e.g. matrix vector multiplication). Perhaps {\em the} example of this is the ubiquitous low rank approximation. Udell and Townsend give a theoretical justification for why low rank approximation is so ubiquitous in machine learning applications~\cite{UT}. 
Now we present a (very incomplete) sampler of other applications of structured matrices in machine learning-- convolutions for image, language, and speech modeling~\cite{gu2018recent}, and low-rank and sparse
matrices for efficient storage and inference on edge devices~\cite{yu2017compressing}.
Forms of structure such as sparsity have been at the forefront of recent advances in machine learning~\cite{frankle2019lottery}, and
are critical for on-device and energy-efficient models, two application areas
of tremendous recent interest~\cite{tsidulko2019google, schwartz2019green}.

At a very high level, the main question we consider in this survey is if there is a similar family of structured matrices that has all the nice properties of low rank approximation but are more expressive than low rank matrices (e.g. many of the transforms including the Fourier transform are full rank). 

\paragraph{Deep learning} is ubiquitous in our daily lives~\cite{DL} with far reaching consequences-- both good and bad\footnote{This has led to deep intellectual research on societal implications of machine learning even in the theory community~\cite{fairness-book}.}. For this survey, we will focus on the mathematical aspects of deep learning since a treatment of the societal implications of deep learning is out of the scope of this survey. Even a broad theoretical study of neural networks (which form the basis of deep learning) is beyond the scope of this survey and there is a lot of excellent literature on this topic~\cite{DL-theory} that we will side-step.

Instead, we will focus on the issue that deep learning models are getting to be too big (which in parallel raises\footnote{OK, we could not resist. This though is the last mention of societal issues in the survey.} its own ethical issues~\cite{s-parrot}). This for example, can be an issue when trying to store these models (and run inference) on mobile platforms like smartphones. In addition, the state-of-the-art language models have so many parameters that creating such models is not possible outside of large technology companies. While there are many reasons for this, {\em one typical reason} is that these neural networks tend to learn {\em un}structured matrices as part of the neural network model (see Section~\ref{sec:nn-toy} for why matrices make an appearance in neural network architectures). Apriori, the advantage of learning from the set of all possible matrices is that it gives the training algorithm the `best' chance to learn the most expressive matrix. However, given that in many situations there is a budget on how many parameters we can use in representing the matrices, the high level question we consider in this survey is:
\begin{ques}
\label{ques:main-informal}
Given a budget on number of parameters that one can use to represent a matrix, what is the `most expressive' family of matrices?
\end{ques}
We remark that a {\em lot} of recent innovations in deep learning have come from designing new architectures of neural networks, which needless to say, is out of scope for the survey (and the author!). In particular, in this survey we will consider a toy version of a {\em single layer} neural network, which by definition is {\em not so deep}.

\paragraph{Organization of the survey.} We present some preliminaries and background before formalizing Question~\ref{ques:main-informal} in Section~\ref{sec:prelim}. We also formalize the problem of training a neural network (the Baur-Strassen theorem~\cite{BS83} plays a starring role) in Section~\ref{sec:gradient}. In Section~\ref{sec:exist}, we analyze existing families of structured matrices and show how they all fall short in answering Question~\ref{ques:main-informal} (or more precisely its formal version Question~\ref{ques:main}). In Section~\ref{sec:butterfly}, we survey results from Dao et al.~\cite{k-mat} who present (to the best of our knowledge) a new family of structured matrices that indeed answers Question~\ref{ques:main} in the affirmative. We conclude with a (biased) list of open questions in Section~\ref{sec:concl}.

\section{Preliminaries and Problem Definition}
\label{sec:prelim}

We begin by setting up notation in Section~\ref{sec:notation}. We setup necessary background in Section~\ref{sec:mvm} (matrix vector multiplication), Section~\ref{sec:nn-toy} (neural networks), Section~\ref{sec:structured} (structured matrices) and Section~\ref{sec:ac} (arithmetic circuits). Finally, we formalize Question~\ref{ques:main-informal} in Section~\ref{sec:prob-def}.

\subsection{Notation}
\label{sec:notation}

We use $\F$ to denote a field\footnote{We will pretty much use $\F=\R$ (real number) or $\F=\C$ (complex numbers) in the survey. Even though most of the results in the survey can be made to work for finite fields, we will ignore this aspect of the results.}. The set of all length $n$ vectors and $m\times n$ matrices over $\F$ are denoted by $\F^n$ and $\F^{m\times n}$ respectively.

We will denote the entry in $\vW\in\F^{m\times n}$ corresponding to the $i$th row and $j$th column as $\vW[i,j]$. The $i$th row of $\vW$ will be denoted by $\vW[i,:]$. Similarly, the $i$th entry in the vector $\vx$ will be denoted as $\vx[i]$. We will follow the  convention that the indices $i$ and $j$ start at $0$. The inner product of vectors $\vx$ and $\vy$ will be denoted by $\ip{\vx}{\vy}$. For any $\vx\in\F^n$, we will use $\diag(\vx)$ to denote the diagonal matrix with $\vx$ being its diagonal.

We will be using asymptotic notation and use $\tO{\cdot}$ to hide poly-log factors in the Big-Oh notation.

\subsection{Matrix Vector Multiplication}
\label{sec:mvm}

We now define the matrix-vector multiplication problem that  will be central to the survey: 
\prob{An $m\times n$ matrix $\vW\in \F^{m\times n}$ and a vector $\vx\in \F^n$ of length $n$}{Their product, which is denoted by
                                \[\vy =\vW\cdot \vx,\]
                        where $\vy\in \F^m$ is  a vector of length $m$ and its $i$th entry for $0\le i< m$ is defined as follows:
                                \[\vy[i] =\sum_{j=0}^{n-1} \vW[i,j]\cdot \vx[j].\]
}

One can easily verify that the naive algorithm that basically operationalizes the above definition takes $O(mn)$ operations in the worst-case. Further, if the matrix $\vW$ is arbitrary, one would need $\Omega(mn)$ time (this follows from a simple adversarial argument). 
Assuming that each operation for an field $\F$ can be done in $O(1)$ time, this implies that the worst-case complexity of matrix-vector multiplication is $\Theta(mn)$.

If we just cared about worst-case complexity, we would be done. However, since there is a fair bit of survey left after this spot it is safe to assume that this is not all we care about. It turns out that in a large number of practical applications, the matrix $\vW$ is fixed (or more appropriately has some structure). Thus, when designing algorithms to compute $\vW\cdot \vx$ (for arbitrary $\vx$), we can exploit the structure of $\vW$ to obtain a complexity that is asymptotically better than $O(mn)$.

Next, we take a brief detour into deep learning and motivate why one would need structured matrices in that application.

\subsection{Neural Networks and (not so) deep learning}
\label{sec:nn-toy}

\note{\textbf{WARNING:} We do not claim to have any non-trivial knowledge (deep or otherwise) of deep learning. Thus, we will only consider a very simplified model of neural networks and our treatment of neural networks should in no way be interpreted as being representative of the current state of deep learning.}

We consider a toy version of neural networks in use today: we will consider the so called {\em single layer} neural network:
\begin{definition}
\label{def:nn}
We define a {\em single layer neural network} with input $\vx\in\F^n$ and output $\vy\in\F^m$ where the output is related to input as follows:
\[\vy=g\left(\vW\cdot \vx\right),\]
where $\vW\in\F^{m\times n}$ and $g:\F^m\to\F^m$ is a non-linear function.
\end{definition}

Some remarks are in order: (1) In practice, neural networks are defined for $\F=\R$ or $\F=\C$; 
 (2) One of the common examples of non-linear function $g:\R^m\to\R^m$ is applying to so called $\relu$ function to each entry.\footnote{More precisely, we have $\relu(x)=\max(0,x)$ for any $x\in\R$ and for any $\vz\ni\R^m$, $g(\vz)=(\relu(\vz[0]),\cdots,\relu(\vz[m-1]))$.} (3) The entries in the matrix $\vW$ are typically called the {\em weights} in the layer.

Neural networks have two tasks associated with it: the first is the task of learning the network. For the network in Definition~\ref{def:nn}, this implies learning the matrix $\vW$ given a set of training data $(\vx_0,\vy_0), (\vx_1,\vy_1),\cdots$ where $\vy_i$ is supposed to be a noisy version of $g(\vW\vx)$-- we will come back to this in Section~\ref{sec:nn-train}. 

The second task is that once we have learned $\vW$, we use it to {\em classify} new data points $\vx$ by computing $g(\vW\vx)$. In practice, we would like the second step to be as efficient as possible.\footnote{Ideally, we would also like the first step to be efficient but typically the learning of the network can be done in an offline step so it can be (relatively) more inefficient.} Ideally we should be able to compute $g(\vW\vx)$ with $O(m+n)$ operations. The computational bottleneck in computing $g(\vW\vx)$ is computing $\vW\cdot \vx$. Further, it turns out (as well will see later in Section~\ref{sec:gradient}) that the complexity of the first step of learning the network is closely related to the complexity of the corresponding matrix-vector multiplication problem.

\subsection{Structured Matrices}
\label{sec:structured}

As mentioned above, in the deep learning setup, we would like to have weight matrices $\vW$ such that the matrix-vector multiplication $\vW\vx$ for an arbitrary $\vx\in\F^n$ can be done in near-linear time. However, if the matrix $\vW$ is represented in the usual $m\times n$ matrix format, then we end up with an $\Omega(mn)$ time just to read the entries of $\vW$. Thus, to have any hope of near-linear matrix vector multiplication, we need to have a smarter representation of the structured matrices. We first recall two examples of structured matrices that have wide applicability in numerical linear algebra and machine learning.


We begin with the notion of low-rank matrices (which are ubiquitous in machine learning~\cite{UT}):
\begin{definition}[Low rank matrices]
\label{def:low-rank}
A matrix $\vW\in \F^{m\times n}$ has rank $r$ (for $0\le r\le \min(m,n)$) if and only if there exists matrices $\vL\in \F^{m\times r}$ and $\vR\in \F^{r\times n}$ such that
\[\vW=\vL\cdot \vR.\]
\end{definition}

It is easy to see that rank $r$ matrices can be represented in $r(n+m)$ elements (by storing $\vL$ and $\vR$) and also has an $O(r(n+m))$-operations matrix vector multiplication by computing $\vW\vx$ as $\paren{\vL\cdot \paren{\vR\cdot \vx}}$. Thus, constant rank matrices indeed satisfy the linear-time matrix-vector multiplication desiderata.

Next, we consider sparse matrices:
\begin{definition}[Sparse matrices]
\label{def:sparse}
A matrix $\vW\in \F^{m\times n}$ is $s$ sparse (for $0\le s\le mn$) if at most $s$ entries in $\vW$ are non-zero.
\end{definition}

The defacto representation of sparse matrices is the {\em listing representation}, where one keeps a list of the locations of the $s$ non-zero values along with the actual non-zero value (every entry not in this list has a value of $0$). Assuming the listing representation, the obvious modification to the naive matrix-vector multiplication (where we automatically `skip' over entries $(i,j)$ such that $\vW[i,j]=0$) results in an $O(s)$ operations algorithm. Thus, $\tO{n}$-sparse matrices indeed satisfy the linear-time matrix-vector multiplication desiderata.

Next, we consider more algebraic families of matrices. 
Consider the {\em discrete Fourier matrix}:
\begin{definition}
\label{def:fourier-mat-complex}
The $n\times n$ {\em discrete Fourier} matrix $\vF_n$ defined as follows (for $0\le i,j<n$):
\[F_{n}[i,j]=\omega_n^{ij},\]
where $\omega_n = e^{-2\pi\iota/n}$ is the $n$-th {\em root of unity} and $\iota=\sqrt{-1}$.
\end{definition}
We note that even though the discrete Fourier matrix has rank $n$ and sparsity $n^2$, it has a very simple representation: just the number $n$.

Let us unroll the following matrix-vector multiplication: $\vhx=\vF_n\vx$. In particular, for any $0\le i<n$:
\[\vhx[i]=\sum_{j=0}^{n-1} \vx[j]\cdot e^{2\pi \iota ji/n}.\]
In other words, $\vhx$ is the {\em discrete Fourier transform} of $\vx$. It turns out that the discrete Fourier transform is incredibly useful in practice (and is used in applications such as image compression). One of the most celebrated algorithmic results is that the Fourier transform can be computed with $O(n\log{n})$ operations:
\begin{theorem}[Fast Fourier Transform (FFT)~\cite{fft}]
\label{thm:chap1:fft}
For any $\vx\in\C^n$, one can compute $\vF_n\cdot \vx$ in $O(n\log{n})$ operations.
\end{theorem}
Thus, the discrete Fourier transform satisfies the near linear-time matrix-vector multiplication desiderata.

Consider the following matrix (called a Vandermonde matrix):
\begin{definition}[Vandermonde Matrix]
\label{def:chap1:vandermonde}
For any $n\ge 1$ and any field $\F$ with size at least $m$, $m$ distinct elements $a_0,\dots,a_{m-1}\in \F$, consider the matrix (where $0\le i <m$ and $0\le j<n$)
\[\vV_n^{(\va)}[i,j]=a_i^j,\]
where $\va=(a_0,\dots,a_{m-1})$.
\end{definition}

We now state some interesting facts about these matrices (which also show that Vandermonde matrices satisfy the near linear-time matrix-vector multiplication desiderata):
\begin{enumerate}
\item One can represent a Vandermonde matrix by noting $a_0\dots,\alpha_{m-1}$ (along with $n$ of course).
\item The discrete Fourier matrix is a special case of a Vandermonde matrix. 
\item The Vandermonde matrix has full rank and has sparsity $n^2$.
\item It turns out that $\vV_n\cdot \vx$ for any $\vx\in\F^n$ can be computed with $O(n\log^2{n})$ operations~\cite{burgisser2013algebraic}. 
\end{enumerate}

In all the four examples of structured matrices that we have seen in this section, their representation pretty much follows from their definitions. However, in general, whenever we have a {\em family} of {\em structured matrices}, we would like a generic way of referring to the representation. To abstract this we will assume that
\begin{assum}
\label{assum:chap4:marix-class}
Given a vector $\vtheta\in\F^{s}$ for some $s=s(m,n)$ such that the vector $\vtheta$ completely specifies a matrix in our chosen family. We will use $\vW_{\vtheta}$ to denote the class of matrix family parameterized by $\vtheta$.
\end{assum}
For example, if $s=mn$, then we get the set of all matrices in $\F^{m\times n}$. On the other hand, for say the Vandermonde matrix (recall Definition~\ref{def:chap1:vandermonde}), we have $s(m,n)=m$ and $\vtheta=(a_0,\dots,a_{m-1})$ for distinct $a_i$'s. 



\subsection{Arithmetic Circuits}
\label{sec:ac}

So far we have tip-toed around how to determine the `optimal' matrix vector multiplication time for a given $\vW$. Now, we pay closer attention to this problem:
\begin{ques}
\label{ques:pipe-dream-gen}
Given an $m\times n$ matrix $\vW$, what is the optimal complexity of computing $\vW\cdot \vx$ (for arbitrary $\vx$)?
\end{ques}

Note that to even begin to answer the question above,
we need to fix our `machine model.'
One natural model is the RAM model on which we analyze most of our beloved algorithms. However, we do not understand the power of RAM model (in the sense that we do not have a good handle on what problems can be solved by say linear-time or quadratic-time algorithms\footnote{The reader might have noticed that we are ignoring the $\P$ vs. $\NP$ elephant in the room.}) and answering Question~\ref{ques:pipe-dream-gen} in the RAM model seems hopeless.

So we need to consider a more restrictive model of computation. Instead of going through a list of possible models, we will just state the model of computation we will use: {\em arithmetic circuit} (also known as the straight-line program). In the context of an arithmetic circuit that computes $\vy=\vW\vx$, there are $n$ inputs gates (corresponding to $\vx[0],\dots,\vx[n-1]$) and $m$ output gates (corresponding to $\vy[0],\dots,\vy[m-1]$). All the internal gates correspond to the addition, multiplication, subtraction and division operators over the underlying field $\F$. The circuit is also allowed to use constants from $\F$ for `free.' The complexity of the circuit will be its {\em size}: i.e. the number of addition, multiplication, subtraction and division gates in the circuit. We will also care about the {\em depth} of the circuit, which is the depth of the DAG representing the circuit. Let us record this choice:

\begin{definition}
\label{def:arith-comp}
For any function $f:\F^n\to \F^m$, its {\em arithmetic circuit complexity} is the minimum number of addition, multiplication, subtraction and division operations over $\F$ needed to compute $f(\vx)$ for any $\vx\in\F^n$.
\end{definition}

Given the above, we have the following more specific version of Question~\ref{ques:pipe-dream-gen}:

\begin{ques}
\label{ques:pipe-dream-arith}
Given a matrix $\vW\in \F^{m\times n}$, what is the arithmetic circuit complexity of computing $\vW\cdot \vx$ (for arbitrary $\vx\in\F^n$)?
\end{ques}

One drawback of arithmetic circuits (especially for infinite fields e.g. $\F=\R$, which is our preferred choice for deep learning applications) is that they assume operations over $\F$ can be performed {\em exactly}. In particular, it ignores precision issues involved with real arithmetic. Nonetheless, this model turns out to be a very useful model in reasoning about the complexity of doing matrix-vector multiplication for any family of matrices.

Perhaps the strongest argument in support of arithmetic circuits is that a large (if not an overwhelming) majority of matrix-vector multiplication algorithm in the RAM model also imply an arithmetic circuit of size comparable to the runtime of the algorithm (and the depth of the circuit roughly corresponds to the time taken to compute it by a parallel algorithm). For example consider the obvious algorithm to compute $\vW\vx$ (i.e. for each $i\in [m]$, compute $\vy[i]$ as the sum $\sum_{i=0}^{n-1} \vW[i,j]\vx[j]$). It is easy to see that this algorithm implies an arithmetic circuit of size $O(nm)$ and depth $O(\log{n})$. 

One reason for the vast majority of existing efficient matrix vector algorithms leading to arithmetic circuits is that they generally are divide and conquer algorithms that use polynomial operations such as polynomial multiplication or evaluation (both of which themselves are divide and conquer algorithms that use FFT (Theorem~\ref{thm:chap1:fft}) as a blackbox) or polynomial addition. Each of these pieces are well known to have small (depth and size) arithmetic circuits (since FFT has these properties). Finally, the divide and conquer structure of the algorithms leads to the circuit being of low depth. See the book of Pan~\cite{pan-book} for a more elaborate description of this connection.

\subsubsection{Linear circuit complexity}

Next, instead of considering  the general arithmetic circuit complexity of $\vW\vx$, let us consider the linear arithmetic circuit complexity. A linear arithmetic circuit only uses linear operations:
\begin{definition}
A linear arithmetic circuit (over $\F$) only allows operations of the form $\alpha X+ \beta Y$, where $\alpha,\beta\in\F$ are constants while $X$ and $Y$ are the inputs to the operation. The {\em linear arithmetic circuit complexity} of $\vW\vx$ is the size of the smallest linear arithmetic circuit that computes $\vW\vx$ (where $\vx$ are the inputs and the circuit depends on $\vW$). Sometimes we will overload terminology and call the (linear) arithmetic circuit complexity of computing $\vW\vx$ as the (linear) arithmetic circuit complexity of (just) $\vW$.
\end{definition}


We first remark that the linear arithmetic circuit complexity seems to be a very natural model to consider the complexity of computing $\vW\vx$ (recall this defines a linear function over $\vx$). In fact one could plausibly conjecture that going from general arithmetic circuit complexity to linear arithmetic circuit complexity of computing $\vW\vx$ should be without loss of generality (the intuition being: "What else can you do?").

It turns out that for infinite fields, the above intuition is correct:
\begin{theorem}[\cite{burgisser2013algebraic}]
\label{thm:line-arith=arith}
Let $\vF$ be an infinite field and $\vW\in\F^{m\times n}$.
Let $\cC(\vW)$ and $\cC^L(\vW)$ be the arithmetic circuit complexity and linear arithmetic circuit complexity of computing $\vW\vx$ (for arbitrary $\vx$). Then $\cC^L(\vW)=\Theta\left(\cC(\vW)\right)$.
\end{theorem}


We first make some observations.
First, it turns out that Theorem~\ref{thm:line-arith=arith} can be proved for finite fields that are exponentially large. 
Second, it is a natural question to try and prove a version of Theorem~\ref{thm:line-arith=arith} for small finite fields (say over $\F_2$). This question is very much open.

\subsection{Problem Definition}
\label{sec:prob-def}

Finally, we have all the pieces in place so that we formally define the problem we are interested in.

Mainly for notational simplicity, we make the following assumption for the rest of the survey:
\begin{assum}
\label{assum:m=n}
Unless stated otherwise, we will consider square matrices, i.e. $m=n$. 
\end{assum}


As mentioned in Section~\ref{sec:nn-toy}, we would like to use a weight matrix $\vW$ such that computing $\vW\vx$ is efficient. In particular, using our choice of measuring algorithmic efficiency by the arithmetic complexity of computing $\vW\vx$, the design problem becomes the following-- can we design neural  networks with weight matrices $\vW$ that are guaranteed to have an arithmetic circuit of size $s$ (for some $s$ that is at most $o(n^2)$)? 
In the rest of the section, we will successively formalize (and specialize) the above intuitive problem statement.

Recall from Section~\ref{sec:nn-toy} that for neural networks, the main bottleneck is to be able to `learn' these weight matrices $\vW$ from the training data (we will formally state the learning problem in Definition~\ref{def:chap4:nn-train-class} but for now we'll keep the definition of training a bit vague). But even before we talk about the efficiency\footnote{Recall that the training problem happens `offline' so we do not need the learning to be say $O(n)$ time but we would like the learning algorithm to be at the worst be polynomial time.} of learning the matrix $\vW$, we note that it is important to be more precise of the representation that the learning algorithm outputs. In particular, even if $\vW$ has an arithmetic circuit of size $s=o(n^2)$, if the learning algorithm outputs the matrix $\vW$ in the usual $n\times n$ matrix format, then we are still stuck with an $\Omega(n^2)$ arithmetic circuit complexity for the learned matrix $\vW$. 

Thus, we want the learning process to not only learn a matrix $\vW$ with arithmetic circuit complexity $s$ but also to learn a representation from which one can easily create a matrix-vector multiplication algorithm with complexity (roughly) $s$. This implies that we first need to identify a {\em class} of structured matrices that can capture matrices with arithmetic circuit complexity of $s$. 
Allowing for the possibility that we might need more than $s$ parameters to index the class of matrices we are after, here is a more formal version of the problem we had stated earlier:
\begin{itemize}
\item A parameter size $s'\ge s$ and a function $f:\F^{s'}\to\F^{n\times n}$ such that
\begin{enumerate}
\item For every matrix $\vW$ with arithmetic circuit complexity at most $s$, there exists a $\vtheta\in\F^{s'}$ such that $f(\vtheta)=\vW$.
\item Given $\vtheta$ one can efficiently compute $f(\vtheta)\cdot \vx$ (here by efficiently we mean with roughly $\tO{s'}$ arithmetic operations).
\item We can efficiently learn the parameter $\vtheta$ that defines $\vW$.
\end{enumerate}
\item The overall goal would be to make $s'$ as close to $s$ as possible-- ideally we want $s'=\tO{s}$.
\end{itemize}

There is an `obvious' family that almost gets us what we want-- just define the parameter $\vtheta$ to encode the circuit computing $\vW\vx$. The problem with this formulation (other than being not an `interesting' definition) is that there is no known efficient way to learn the optimal arithmetic circuit  for $\vW$ (even if we were given access to the $n\times n$ representation of $\vW$).

Another candidate for the class of circuits we are looking for will be the family of low rank matrices. In particular, given the target $s$, we would like to figure out the value of rank $r$ so that we can pick $s'=rn$ and we use the standard representation of rank $r$ matrices. In this case, it is easy to verify that all the three properties above are satisfied. The problem of learning the rank $r$ decomposition of a given matrix $\vW$ e.g. can be computed by the Singular Value Decomposition (or SVD).\footnote{In fact the SVD will give the best rank $r$ approximation even if $\vW$ is not rank $r$-- for now let's just consider the problem setting where we are looking for an {\em exact} representation.} Unfortunately, in general $s'$ can be much larger than $s$-- consider e.g. the DFT (Definition~\ref{def:fourier-mat-complex}), which has $s=O(n\log{n})$, but since the matrix is full rank, we need $r=n$ and hence $s'=n^2$, which is not that useful.

We will consider some other choices for families of structured matrices in Section~\ref{sec:exist} but before we finalize the problem statement, we use the following observation from practice to make the problem a bit more tractable-- it turns out in practice that the weight matrix $\vW$ (or its representation $\vtheta$) is learned via gradient descent (see Algorithm~\ref{alg:chap4:GD}). So we make the following assumption:
\begin{assum}
\label{assum:GD}
We will assume that we can only use gradient descent to learn the representation $\vtheta$ for our target matrix $\vW$.
\end{assum}
What the above means is that it is {\em sufficient} to be able to compute the gradient of $f$ at any point in $\F^{s'}$ (see Section~\ref{sec:gradient} for details on why this is the case). Under this assumption, we can modify our earlier goal into our final problem statement:

\begin{ques}
\label{ques:main}
Does there exist a family of $n\times n$ matrices such that for every parameter $n\le s\le n^2$, there exists a parameter $s'$ and a map $f:\F^{s'}\to\F^{n\times n}$ such that for every matrix $\vW$ with arithmetic circuit complexity of at most $s$, there exists $\vtheta\in\F^{s'}$ such that $f(\vtheta)=\vW$. Furthermore, we want
\begin{itemize}
\item (\propone) $s'$ is as close to $s$ as possible (ideally $s'=\tO{s}$)
\item (\proptwo) Given $\vtheta$, we can compute $f(\vtheta)\cdot \vx$ for any $\vx\in\F^n$ in close to $s'$ arithmetic operations.
\item (\propthree) For any $\va\in\F^{s'}$, one can evaluate the gradient of $f$ at $\va$ efficiently (ideally as close to $s'$ arithmetic operations as possible).
\end{itemize}
\end{ques}

Before we attack Question~\ref{ques:main}, we will take a bit of a detour to consider the problem of learning $\vW$ from training data in more detail.

\section{Computing gradients}
\label{sec:gradient}

We will formalize the problem of learning from training data in Section~\ref{sec:nn-train}. Then in Section~\ref{sec:grad-suff}, we identify a specific gradient function that is sufficient to run gradient descent for our purposes. We recall the Baur-Strassen theorem in Section~\ref{sec:chap4:auto-diff}, which will show that for our gradient problem, it is enough to ensure that $\vW$ has small arithmetic circuit complexity. Finally, in Section~\ref{sec:transpose}, we take a detour to highlight a really cool result, which unfortunately does not seem to be as well-known as it should be.

We will {\em not} be assuming Assumption~\ref{assum:m=n} in this section, i.e. in this section we will consider a general rectangular matrix $\vW$ (and we will revert to  Assumption~\ref{assum:m=n} from next section onwards).

\subsection{Back to (not so) deep learning}
\label{sec:nn-train}

We go back to the single layer neural network that we studied earlier in Section~\ref{sec:nn-toy}. In particular, recall we consider a single layer neural network that is defined by
\begin{equation}
\label{eq:chap4:nn}
\vy=g\left(\vW\cdot \vx\right),
\end{equation}
where $\vW\in\F^{m\times n}$ and $g:\F^m\to\F^m$ is a non-linear function. Further,
\begin{assum}
\label{assum:chap4:non-lin}
We will assume that non-linear function  $g:\F^m\to\F^m$ is obtained by applying the same function $g:\F\to\F$ to each of the $m$ elements.
\end{assum}
In other words,~\eqref{eq:chap4:nn} is equivalently stated as for every $0\le i<m$:
\[\vy[i]= g\paren{\ip{\vW[i,:]}{\vx}}.\]

Recall that in Section~\ref{sec:nn-toy}, we had claimed (without any argument) that the complexity of learning the weight matrix $\vW$ given few samples is governed by the complexity of matrix-vector multiplication for $\vW$. In this section, we will rigorously argue this claim. To do this, we define the learning problem more formally:
\begin{definition}
\label{def:chap4:nn-train-gen}
Given $L$ {\em training data} $\paren{\vx^{(\ell)}, \vy^{(\ell)}}$ for $\ell\in [L]$, we want to compute a matrix $\vW\in\F^{m\times n}$ that minimizes the error
\[E(\vW)=\sum_{\ell=1}^L \norm{2}{\vy^{(\ell)}- g\paren{\vW\cdot \vx^{(\ell)}}}^2.\]
\end{definition}
We note that the above is not the only {\em error function} that is used in training neural networks but the above is a common choice and hence, we stick with it.
Further, note that in the above the training searches for the 'best' weight matrix from the set of all matrices in $\F^{m\times n}$. However, since we are interested in searching for the best weight matrix with a certain class as in Question~\ref{ques:main}, 
we generalize Definition~\ref{def:chap4:nn-train-gen} as follows:
\begin{definition}
\label{def:chap4:nn-train-class}
Given $L$ {\em training data} $\paren{\vx^{(\ell)},\vy^{(\ell)}}$ for $\ell\in [L]$, we want to compute the parameters of an $m\times n$ matrix $\vtheta\in\F^{s(m, n)}$ that minimizes the error (where we use $\vW_\vtheta=f(\vtheta)$):
\[E(\vtheta)=\sum_{\ell=1}^L \norm{2}{\vy^{(\ell)}- g\paren{\vW_{\vtheta}\cdot \vx^{(\ell)}}}^2.\]
\end{definition}

\subsubsection{Gradients and Gradient Descent}

\paragraph{(Partial) Derivatives.} It turns out that we will only be concerned with studying derivatives of polynomials. For this, we can define the notion of a formal derivative (over univariate polynomials):
\begin{definition}
\label{def:deriv}
The formal derivative $\deriv{X}{\cdot}:\F[X]\to\F[X]$ is defined as follows. For every integer $i$,
\[\deriv{X}{X^i}= i\cdot X^{i-1}.\]
The above definition can be extended to all polynomials in $\F[X]$ by insisting that $\deriv{X}{\cdot}$ be a linear map. That is for every $\alpha,\beta\in\F$ and $f(X),g(X)\in \F[X]$ we have
\[\deriv{X}{\alpha f(X)+\beta g(X)} = \alpha\deriv{X}{f(X)}+\beta\deriv{X}{g(X)}.\]
\end{definition}
We note that over $\R$, the above definition when applied to polynomials over $\R[X]$ gives the same result as the usual notion of derivatives.

We will actually need to work with derivatives of multi-variate polynomials. We will use $\F[X_1,\dots,X_m]$ to denote the set of multivariate polynomials with variables $X_1,\dots,X_m$. For example, $3XY+Y^2+1.5 X^3Y^4$ is in $\R[X,Y]$. We extend the definition of derivatives from Definition~\ref{def:deriv} to the following (which also called a {\em gradient})
\begin{definition}
\label{def:gradient}
Let $f(X_1,\dots,X_n)$ be a polynomial in $\F[X_1,\dots,X_n]$. Then define its {\em gradient} as (where we use $\vX=(X_1,\dots,X_n)$ to denote the vector of variables):
\[\deriv{\vX}{f(\vX)}=\paren{\deriv{X_1}{f(\vX)},\dots,\deriv{X_n}{f(\vX)}},\]
where in $\deriv{X_i}{f(\vX)}$, we think of $f(\vX)$ as being a polynomial in $X_i$ with coefficients in $\F[X_1,\dots,X_{i-1},X_{i+1},\dots, X_n]$.

Finally note that $\deriv{X_i}{f(\vX)}$ is again a polynomial and we will denote its evaluation at $\va\in\F^n$ as $\deriv{X_i}{f(\vX)}_{|\va}$. We extend this notation to the gradient by
\[\deriv{\vX}{f(\vX)}_{|\va}=\paren{\deriv{X_1}{f(\vX)}_{|\va},\dots,\deriv{X_n}{f(\vX)}_{|\va}}.\]
\end{definition}
For example
\[\deriv{X,Y}{3XY+Y^2+1.5 X^3Y^4}=\paren{3Y+4.5X^2Y^4,3X+2Y+6X^3Y^3}.\]


\paragraph{Gradient Descent.} While there exist techniques to solve the above problem theoretically, in practice {\em Gradient Descent} is commonly used to solve the above problem. In particular, one starts off with an initial state $\vtheta=\vtheta_0\in\F^s$ and one keeps changing $\vtheta$ is opposite direction of $\deriv{\vtheta}{E(\vtheta)}$ till the error is below a pre-specified threshold (or one goes beyond a pre-specified number of iterations). Algorithm~\ref{alg:chap4:GD} has the details.

\begin{algorithm}
\caption{Gradient Descent}
\label{alg:chap4:GD}
\begin{algorithmic}[1]
\Require{$\eta>0$ and $\eps>0$}
\Ensure{$\vtheta$}
\Statex
\State $i\gets 0$
\State Pick $\vtheta_0$ \Comment{This could be arbitrary or initialized to something more specific}
\While{$|E(\vtheta_i)|\ge \eps$} \Comment{One could also terminate based on number of iterations}
    \State $\vtheta_{i+1}\gets \vtheta_i -\eta\cdot\paren{\deriv{\vtheta}{E(\vtheta)}}_{|\vtheta_i}$ \Comment{$\eta$ is the 'learning rate'}
    \State $i\gets i+1$
\EndWhile
\State \Return{$\vtheta_i$}
\end{algorithmic}
\end{algorithm}

\subsection{Computing the gradient}
\label{sec:grad-suff}

It is clear from Algorithm~\ref{alg:chap4:GD}, that the most computationally intensive part is computing the gradient. We first show that if one can compute a related gradient, then we could implement Algorithm~\ref{alg:chap4:GD}. In Section~\ref{sec:chap4:auto-diff} we will show that this latter gradient computation is closely tied to computing $\vW\vx$. We first argue:
\begin{lemma}
\label{lem:chap4:related-grad}
If for every $\vz\in\F^m$ and $\vu\in\F^n$, one can compute $\paren{\deriv{\vtheta}{\vz^T\vW_{\vtheta}\vu}}_{|\va}$ for any $\va\in\F^s$ in $T_1(m,n)$ operations and $\vW\vu$ in $T_2(m,n)$ operations, then one can compute  $\paren{\deriv{\vtheta}{E(\vtheta)}}_{|\vtheta_0}$ for a fixed $\vtheta_0\in\F^s$ in $O(L(T_1(m,n)+T_2(m,n)))$ operations.
\end{lemma}
\begin{proof}
For notational simplicity define
\[\vW=\vW_{\vtheta_0}\]
and
\[E_{\ell}(\vtheta)= \norm{2}{\vy^{(\ell)}- g\paren{\vW_{\vtheta}\cdot \vx^{(\ell)}}}^2.\]
Fix $\ell\in [L]$.
We will show that we can compute $\deriv{\vtheta}{E_{\ell}(\vtheta)}_{|\vtheta_0}$ with $O(T_1(m,n)+T_2(m,n))$ operations, which would be enough since $\deriv{\vtheta}{E(\vtheta)}=\sum_{\ell=1}^L \deriv{\vtheta}{E_{\ell}(\vtheta)}$.

For notational simplicity, we will use $\vy,\vx$ and $E(\vtheta)$ to denote $\vy^{(\ell)},\vx^{(\ell)}$ and $E_{\ell}(\vtheta)$ respectively. Note that
\begin{align*}
E(\vtheta) &= \norm{2}{\vy- g\paren{\vW_{\vtheta}\cdot \vx}}^2\\
&=\sum_{i=0}^{m-1}\paren{\vy[i]-g\paren{\sum_{j=0}^{n-1}\vW_{\vtheta}[i,j]\vx[j]}}^2.
\end{align*}
Applying the chain rule of the gradient on the above, we get (where $g'(x)$ is the derivative of $g(x)$):
\begin{equation}
\label{eq:E-grad}
\deriv{\vtheta}{E(\vtheta)} = -2\sum_{i=0}^{m-1}\paren{\vy[i]-g\paren{\sum_{j=0}^{n-1}\vW_{\vtheta}[i,j]\vx[j]}}g'\paren{\sum_{j=0}^{n-1}\vW_{\vtheta}[i,j]\vx[j]}\sum_{j=1}^{n-1}\paren{\deriv{\vtheta}{\vW_{\vtheta}[i,j]}\vx[j]}.
\end{equation}
Define a vector $\vz\in\F^m$ such that for any $0\le i<m$,
\[\vz[i]=-2\paren{\vy[i]-g\paren{\ip{\vW[i,:]}{\vx}}}g'\paren{\ip{\vW[i,:]}{\vx}}.\]
Note that once we compute $\vW\vx$ (which by assumption we can do in $T_2(m,n)$ operation), we can compute $\vz$ with $O(T_2(m,n))$ operations.\footnote{Here we have assumed that one can compute $g(x)$ and $g'(x)$ with $O(1)$ operations and assumed that $T_2(m,n)\ge m$.} Further, note that $\vz$ is independent of $\vtheta$ (recall $\vW=\vW_{\vtheta_0}$).

From~(\ref{eq:E-grad}), we get that
\begin{align*}
\deriv{\vtheta}{E(\vtheta)}_{|\vtheta_0}&= -2\sum_{i=0}^{m-1} \paren{\vy[i]-g\paren{\ip{\vW[i,:]}{\vx}}}g'\paren{\ip{\vW[i,:]}{\vx}}\sum_{j=0}^{n-1} \paren{\deriv{\vtheta}{\vW_{\vtheta}[i,j]}_{|\vtheta_0}\cdot \vx[j]}\\
&= \sum_{i=0}^{m-1} \vz[i]\cdot \sum_{j=0}^{n-1} \paren{\deriv{\vtheta}{\vW_{\vtheta}[i,j]}_{|\vtheta_0}\cdot \vx[j]}\\
&= \paren{\deriv{\vtheta}{\sum_{i=0}^{m-1} \vz[i]\cdot \sum_{j=0}^{n-1} {\vW_{\vtheta}[i,j]}\cdot \vx[j]}}_{|\vtheta_0}\\
&=\paren{\deriv{\vtheta}{\vz^T\vW_{\vtheta}\vx}}_{|\vtheta_0}.
\end{align*}


In the above, the first equality follows from our notation that $\vW=\vW_{\vtheta_0}$, the second equality follows from the definition of $\vz$ and the third equality follows from the fact that $\vz$ is independent of $\vtheta$. The proof is complete by noting that we can compute $\paren{\deriv{\vtheta}{\vz^T\vW_{\vtheta}\vx}}_{|\vtheta_0}$ in $T_1(m,n)$ operations.
\end{proof}

Thus, to efficiently implement gradient descent, we have to efficiently compute  $\paren{\deriv{\vtheta}{\vz^T\vW_{\vtheta}\vx}}_{|\vtheta_0}$ for any fixed $\vz\in\F^m$ and $\vx\in\F^n$. Next, we will show that the arithmetic complexity of this operation is the same (up to constant factors) as the arithmetic complexity of computing $\vz^T\vW\vx$ (which in turn has complexity no worse than that of computing our old friend $\vW\vx$). In the next section, not only will we show that this result is true but it is true for {\em any} function $f:\F^s\to\F$. As a bonus, we will present a simple (but somewhat non-obvious) algorithmic proof.

\subsection{Computing gradients very fast}
\label{sec:chap4:auto-diff}

In this section we consider the following general problem:
\prob{An arithmetic circuit $\cC$ that computes a function $f:\F^s\to\F$ and an evaluation point $\va\in\F^s$.}{$\deriv{\vtheta}{f(\vtheta)}_{|\va}$.}

Recall that in the previous section, we were interested in solving the above problem for the function $f_{\vz,\vx}(\vtheta)=\vz^T\vW_{\vtheta}\vx$ where $\vW_{\vtheta}\in\F^{m\times n},\vz\in\F^m$ and $\vx\in\F^n$.

The way we will tackle the above problem is given the arithmetic circuit $\cC$ for $f(\vtheta)$, we will try to come up with an arithmetic circuit $\cC'$ to compute $\deriv{\vtheta}{f(\vtheta)}$. We first note that given a fixed $0\le \ell<s$, it is fairly easy compute a circuit $\cC'_{\ell}$ that on input $\va\in\F^s$ computes $\deriv{\vtheta[\ell]}{f(\vtheta)}_{|\va}$ with essentially the same size. 
This implies that one can compute $\deriv{\vtheta}{f(\vtheta)}$ with arithmetic circuit complexity $O(m\cdot|\cC|)$ (where $|\cC|$ denotes the size of $\cC$).

We will now recall the Baur-Strassen theorem, which states that the gradient can be computed in the same (up to constant factors) arithmetic circuit complexity as evaluating $f$.

\begin{theorem}[Baur-Strassen Theorem~\cite{BS83}]
\label{thm:chap4:baur-strassen}
Let $f:\F^s\to\F$ be a function that has an arithmetic circuit $\cC$ such that given $\vtheta\in\F^s$, it computes $f(\vtheta)$. Then there exists another arithmetic circuit $\cC'$ that computes for any given $\va\in\F^s$, the gradient $\deriv{\vtheta}{f(\vtheta)}_{|\va}$. Further,
\[|\cC'|\le O(|\cC|).\]
\end{theorem}

The proof of Baur-Strassen theorem is actually algorithmic-- Algorithm~\ref{alg:chap4:back-prop} shows how to compute the gradient given the arithmetic circuit for $f$ (it is not too hard to see that the algorithm implicitly defines the claimed arithmetic circuit $\cC'$). The proof of correctness of the algorithm follows from
%
 the following version of chain rule for multi-variable function.
\begin{lemma}
\label{lem:chap4:chain-rule-multi}
Let $f:\F^s\to\F$ be a function composition of a polynomial $g\in \F[H_1,\dots,H_k]$ and polynomials $h_i\in \F[X_1,\dots,X_s]$ for every $i\in [k]$, i.e.
\[f(\vX)=g\paren{h_1(\vX),\dots,h_k(\vX)}.\]
Then for every $0\le \ell <s$, we have
\[\deriv{X_{\ell}}{f(\vX)}=\sum_{j=1}^k \deriv{H_{j}}{g(H_1,\dots,H_k)}\cdot\deriv{X_{\ell}}{h_{j}(\vX)}.\]
\end{lemma}
We note that over $\R$ the above is known as the {\em high-dimensional chain rule} (and it holds for more general classes of functions). It turns out that if $g$ and $h_i$ are polynomials, then the high-dimensional chain rule pretty much follows from Definition~\ref{def:deriv}. 

\begin{algorithm}
\caption{Back-propagation Algorithm}
\label{alg:chap4:back-prop}
\begin{algorithmic}[1]
\Require{$\cC$ that computes a function $f:\F^s\to\F$ and an evaluation point $\va\in\F^s$}
\Ensure{$\deriv{\vtheta}{f(\vtheta)}_{|\va}$}
\Statex
\State Let $\sigma$ be an ordering of gates of $\cC$ in reverse topological sort with output gate first \Comment{This is possible since the graph of $\cC$ is a DAG}
\While{Next gate $g$ in $\sigma$ has not been considered}
    \State Let the parent gates of $g$ be $h_1,\dots,h_k$\Comment{$k=0$ is allowed and implies no parents}
    \If{$k=0$}
        \State $\vd[g]\gets 1$
    \Else
        \State $\vd[g]\gets 0$
        \For{$i\in [k]$}
            \State $\vd[g]\gets \vd[g]+\deriv{g}{h_i}_{|\va}\cdot \vd[h_i]$ 
        \EndFor
    \EndIf
\EndWhile
\State \Return{$\paren{\vd[\theta_i]}_{0\le i<s}$} \Comment{$\theta_0,\dots,\theta_{s-1}$ are input gates}
\end{algorithmic}
\end{algorithm}

Theorem~\ref{thm:chap4:baur-strassen} and Lemma~\ref{lem:chap4:related-grad} imply the following connection between the gradient we want to compute the arithmetic circuit complexity of the corresponding matrix-vector multiplication problem:
\begin{corollary}
\label{cor:connect-grad-mvm}
If for every $\vtheta\in\F^s$, $\vW_\vtheta$ has arithmetic circuit complexity of $m$, then we can compute $\paren{\deriv{\vtheta}{E(\vtheta)}}_{|\vtheta_0}$ for  every $\vtheta_0\in\F^s$ in $O(L(m+n))$ operations.
\end{corollary}

\subsubsection{Automatic Differentiation}

It turns out that Algorithm~\ref{alg:chap4:back-prop} can be extended to work beyond arithmetic circuits (at least over $\R$). This uses that fact that the high dimensional chain rule (Lemma~\ref{lem:chap4:chain-rule-multi}) holds for any differentiable functions $g,h_1,\dots,h_k$. In other words, we can consider circuits that compute $f$ where each gate computes a differentiable function of its input. In other words, given a circuit for $f$ with `reasonable' gates, one can automatically compile another circuit for its gradient. This idea has lead to the creation of the field of {\em automatic differentiation} (or {\em auto diff}) and is at the heart of many recent machine learning progress. In particular, those familiar with neural networks would notice that Algorithm~\ref{alg:chap4:back-prop} is the well-known {\em backpropagation algorithm} (and hence the title of Algorithm~\ref{alg:chap4:back-prop}). However, for this survey, we will not need the full power of auto diff (Corollary~\ref{cor:connect-grad-mvm} is all we need).

Next, we take a (wide) detour and state a result that is not as well-known as it should be.
\subsection{Multiplying by the transpose}
\label{sec:transpose}
We first recall the definition of the transpose of a matrix:
\begin{definition}
\label{def:chap4:tranpose}
The {\em transpose} of a matrix $\vA\in\F^{m\times n}$, denoted by $\vA^T\in\F^{n\times m}$ is defined as follows (for any $0\le i<n, 0\le j<m$:
\[\vA^T[i,j]=\vA[j,i].\]
\end{definition}

It is natural to ask (since the transpose it so closely related to the original matrix):
\begin{ques}
\label{ques:chap4:tranpose}
Is the (arithmetic circuit) complexity of computing $\vA^T\vx$ related to the (arithmetic circuit) complexity of computing $\vA\vx$ for {\em every} matrix $\vA\in\F^{m\times n}$? E.g. are they within $\tO{1}$ of each other?
\end{ques}
We will address the above question in the rest of this section. 

\subsubsection{Transposition principle}

It turns out that the answer to Question~\ref{ques:chap4:tranpose} is an emphatic {\em yes}:

\begin{theorem}[Transposition Principle~\cite{1973-fiduccia-matrix}]
\label{thm:chap4:transposition}
Fix a matrix $\vA\in\F^{n\times n}$ such that there exists an arithmetic circuit of size $s$ that computes $\vA\vx$ for arbitrary $\vx\in\F^n$. Then there exists an arithmetic circuit of size $O(s+n)$ that computes $\vA^T\vy$ for arbitrary $\vy\in\F^n$.
\end{theorem}

\note{The above result was surprising to the author when he first came to know about it. Indeed, the knowledge of this result would have saved the author  more than a year's worth of plodding while working on the paper~\cite{soda18}. For whatever reason, this result is not as well-known.} 

It is not too hard to show that the additive $n$ term in the bound in the transposition principle is necessary. 

There exist proofs of the transposition principle that are very structural in the sense that they consider the circuit for computing $\vA\vx$ and then directly change it to compute a circuit for $\vA^T\vy$.\footnote{At a very high level this involves `reversing' the direction of the edges in the DAG corresponding to the circuit.} For this survey we will present a much slicker proof that directly uses the Baur-Strassen theorem (to the best of our knowledge this proof was first explicitly stated in~\cite{transposition}). 
For this the following alternate view of $\vA^T\vy$ will be very useful: 
\begin{equation}
\label{eq:chap4:transpose-equiv}
\vy^T\vA =\paren{\vA^T\vy}^T.
\end{equation}

\begin{proof}[Proof of Theorem~\ref{thm:chap4:transposition}]
Thanks to~(\ref{eq:chap4:transpose-equiv}), we will consider the computation of $\vy^T\vA$ for any $\vy\in\F^n$. We first claim that: 
\begin{equation}
\label{eq:chap4:transposition-gradient}
\vy^T\vA =\deriv{\vx}{\vy^T\vA\vx}.
\end{equation}
Note that the function $\vy^T\vA\vx$ is exactly the same product we have encountered before in Lemma~\ref{lem:chap4:related-grad}.\footnote{However, earlier we where taking the gradient with respect to (essentially) $\vA$ whereas here it is with respect to $\vx$.} Then note that given an arithmetic circuit of size $s$ to compute $\vA\vx$ one can design an arithmetic circuit that computes $\vy^T\vA\vx$ of size $s+O(n)$ (by simply additionally computing $\ip{\vy}{\vA\vx}$, which takes $O(n)$ operations.).

Now, by Theorem~\ref{thm:chap4:baur-strassen}, there is a circuit that computes $\deriv{\vx}{\vy^T\vA\vx}$ with arithmetic circuit of size $O(s+n)$.\footnote{Here we consider $\vA$ as given and $\vx$ and $\vy$ as inputs. This implies that we need to prove the Baur-Strassen theorem when we only take derivatives with respect to part of the inputs-- but this follows trivially since one can just read off $\deriv{\vx}{\vy^T\vA\vx}$ from $\deriv{\vx,\vy}{\vy^T\vA\vx}$.} Equation~(\ref{eq:chap4:transposition-gradient}) completes the proof.
\end{proof}

\section{Towards answering Question~\ref{ques:main}}
\label{sec:exist}

In this section, we walk through some well studied classes of structured matrices and see how they all fall short of answering Question~\ref{ques:main} fully.

\subsection{Low rank matrices}

We start with low rank matrices: we already addressed why low rank matrices cannot be the answer for Question~\ref{ques:main} in Section~\ref{sec:prob-def} but we'll walk through the three requirements again. We consider the standard representation of a rank $r$ matrix $\vW$ as $\vW=\vL\cdot\vR$ for $\vL\in \F^{n\times r}$ and $\vR\in \F^{r\times n}$. In this case $s'=2rn$ and $\vtheta$ is just the listing of all the entries in $\vL$ and $\vR$ and $f$ is defined in the obvious way.

\begin{enumerate}
\item (\propone) We have $s'=2rn$. Consider the case e.g. when $\vW$ is the discrete Fourier matrix, which has rank $r=n$ (and hence $s'\ge \Omega(n^2)$) and by Theorem~\ref{thm:chap1:fft}, $\vW$ has $s=O(n\log{n})$. Thus, \propone\ is not satisfied since the gap between $s'$ and $s$ is pretty much as large as possible.
\item (\proptwo) This property is satisfied since the obvious matrix-vector multiplication algorithm (given $\vL$ and $\vR$) takes $O(rn)$ operations.
\item (\propthree) It is easy to see that each entry in $\vL\cdot \vR$ is a degree two polynomial in the entries of $\vtheta$ and hence is also differentiable.
\end{enumerate}

\subsection{Sparse matrices (in listing representation)} 

Next, we consider $m$ sparse matrices in {\em listing representation}. In other words, $s'=O(m)$ and $\vtheta$ is basically a list of triples $(x_i,y_i,c_i)$ for $1\le i\le m$. The map $f$ is defined as follows:
\[f(\vtheta)[j,k]=
\begin{cases}
c &\text{ if } (j,k,c) \text{ is in } \vtheta\\
0 &\text{ otherwise }
\end{cases}.\]

It turns out that sparse matrices do not satisfy two of the three requirements in Question~\ref{ques:main}--
\begin{itemize}
\item (\propone) We have $s'=\Theta(m)$. However, for the discrete Fourier transform we have $m=n^2$ and as we have already observed that for the discrete Fourier transform we have $s=O(n\log{n})$. Hence, the gap between $s'$ and $s$ is as large as possible.
\item (\proptwo) The obvious algorithm to multiply an $m$-sparse matrix with an arbitrary vector takes $O(m)$ operations and hence \proptwo\ is satisfied.
\item (\propthree) It is easy to check that $f$ as defined above is not differentiable (because the locations of the non-zero values are discrete). E.g. consider the case of $m=1$ and let $(x,y)$ be the location of the non-zero value (and let us assume that $\vW[x,y]=1$). In this case $f(\vtheta)[j,k]=\delta_{x=j,y=k}$, where $\delta$ is the Kronecker delta function for which the derivative is not defined at the point $(x,y,1)$ and hence $f$ is not differentiable.\footnote{In this survey we are dealing with the classical definition of derivatives. If one defines the Kronecker delta function as a limit of a distribution and consider derivatives in the sense of theory of distributions then \propthree\ will be satisfied. Indeed, many practical implementation that use sparse as the weight matrices $\vW$, when trying to learn $\vW$ use the distributional definition of the Kronecker delta function.}
\end{itemize}

As bit of a spoiler alert, (variants) of sparse matrices will actually be crucial in answering Question~\ref{ques:main} in the affirmative. It turns out that to satisfy \propone\ one needs to consider {\em product} of sparse matrices (see Section~\ref{sec:spw}) and to satisfy \propthree\ one needs to go beyond the listing representation (see Section~\ref{sec:butterfly}).

\subsection{Sparse+low rank}
\label{sec:mat-rig}

Next, we consider the combination of sparse and low rank matrices. Not only is this a natural combination to consider but such matrices have been well-studied in the context of {\em robust PCA}~\cite{robust-pca}. However, for this survey we are interested in this family of matrices since this is {\em exactly} the class of matrices considered in the {\em matrix rigidity} problem introduced by Valiant\cite{Val77}. In particular, we recall the following result due to Valiant (where the specific statement is from Paturi and Pudl\'{a}k~\cite{PP06}):

\begin{theorem}[\cite{Val77,PP06}]
\label{thm:matrix-rigidity}
Let $r,d,\sigma$ be positive integers such that $d>4\log_2{\sigma}$. Assume $\vW$ has a circuit $C$ with size
\[s\le r\cdot \frac{\log_2{d}}{2\log_2\paren{\frac d{4\log_2{\sigma}}}},\]
and depth $d$. Then we can decompose $\vW$ as
\[\vW=\vS+\vL\vR,\]
where both $\vS\in \F^{m\times n}$ and $\vR\in\F^{r\times n}$ are $\sigma$- row sparse (i.e. overall they are $m\sigma$ and  $r\sigma$ sparse respectively) and $\vL\in \F^{n\times r}$. In other words, $\vW$ can be written as a sum of rank $r$ and $\sigma n$-sparse matrix.
\end{theorem}

The above result has spawned a long line of beautiful work in the area of matrix rigidity, which we do not have the space to do any justice, see the course notes by Golovnev~\cite{matrix-rig} for more details.

Unfortunately, sparse+low-rank matrices cannot answer Question~\ref{ques:main} positively either. Specifically, we will use the following lower bound result.
\begin{theorem}[Thm 2.17 in~\cite{Lokam09}]
\label{thm:vand-lb}
Let $\va\in\mathbb{Q}^n$ such that all entries in $\va$ are algebraically independent over $\mathbb{Q}$. Then there exists an $\eps$ such that for every $r\le \eps\sqrt{n}$ such that one can write
\begin{equation}
\label{eq:vand-sparse-low-rnk}
\vV_n^{(\va)} = \vS+\vR,
\end{equation}
where $\vR$ has rank $r$, then $\vS$ has overall sparsity at least $\frac{n^2}4$.
\end{theorem}

Let us consider all three required properties in sequence:

\begin{itemize}
\item (\propone) It turns out that the Vandermonde matrices (with the condition as in Theorem~\ref{thm:vand-lb}) still shows that this property is not satisfied for sparse+low rank matrices though the gap between $s'$ and $s$ is not as dramatic as before.  Specifically, we claim that Theorem~\ref{thm:vand-lb} shows\footnote{Indeed consider any sparse+low rank as in~\eqref{eq:vand-sparse-low-rnk}. If $\vR$ has rank $r$ at least $\eps\sqrt{n}$, this immediately implies $s'\ge 2rn \ge \Omega\paren{n^{3/2}}$. If on the other hand if $r\le \eps\sqrt{n}$, then by Theorem~\ref{thm:vand-lb}, we have $s'\ge \frac{n^2}4$.} that $s'\ge \Omega\paren{n^{3/2}}$ (while $s=O(n\log^2{n})$~\cite{burgisser2013algebraic}).
Thus, while the gap is not quadratic as it was for the sparse only or low-rank only case, the gap is still too large for what we are after.
\item (\proptwo) Since this property is satisfied for rank $r$ and $\sigma n$-sparse matrices, this property is also satisfied for their sum.
\item (\propthree) Since this property is not satisfied for sparse matrices (with the listing representation), this property is not satisfied for sum of low rank and sparse matrices as well.
\end{itemize}

We would like to stress that the goal of matrix rigidity is different from ours in that the goal of the program of matrix rigidity is to exhibit an explicit matrix for which any decomposition as $\vR+\vS$ for $\vR$ being rank $O\paren{\frac n{\log\log{n}}}$ needs $\vS$ to have sparsity $\Omega\paren{n^{1+\eps}}$ for some constant $\eps>0$. In our context we would have liked to show that matrices with small arithmetic circuits are {\em not} rigid.

\subsection{Vandermonde matrices}

So far we have been able to rule out low rank, sparse and sparse+low rank matrices just based on the discrete Fourier transform. However, the discrete Fourier transform by itself does not need a lot of parameters. In particular, it is a special case of Vandermonde matrices (Definition~\ref{def:chap1:vandermonde}). It is natural to consider Vandermonde matrices as a potential answer to Question~\ref{ques:main}.
In this case we use the obvious representation where $\vtheta$ is just the vector $(a_1,\dots,a_n)$ and $f$ is defined as per Definition~\ref{def:chap1:vandermonde}. Unfortunately, Vandermonde matrices cannot answer Question~\ref{ques:main} positively either:
\begin{enumerate}
\item (\propone) We have $s'=O(n\log^2{n})$~\cite{burgisser2013algebraic}. However, by a simple counting argument it is easy to see that Vandermonde matrices cannot represent all matrices. Specifically, consider the set of $\bar{s}$-sparse matrices with sparsity $\bar{s}=\omega(n)$. Since a Vandermonde matrix is represented by $n$ parameters, there will be at least one $\bar{s}$-sparse matrix that cannot be represented as a Vandermonde matrix.
Thus, \propone\ is not satisfied. 
\item (\proptwo) This property is satisfied since one can multiply a Vandermonde matrix with an arbitrary vector in $O(n\log^2{n})=O(s')$ operations~\cite{burgisser2013algebraic}.
\item (\propthree) By definition, each entry in a Vandermonde matrix is a polynomial (of degree at most $n-1$) in the entries of $\vtheta$ and hence is also differentiable.
\end{enumerate}

\subsection{Low-displacement rank matrices}
\label{sec:ldr}

We now consider a class of structured matrices that have been used in experiments in deep learning to address the practical questions that motivated Question~\ref{ques:main}.

We begin with the definition of a matrix having a displacement rank of $r$:
\begin{definition}
\label{def:chap5:LDR}
A matrix $\vW\in\F^{n\times n}$ has a {\em displacement rank} with respect to $\vL,\vR\in\F^{n\times n}$, if the {\em residual}
\[\vE=\vL\vW-\vW\vR\]
has rank $r$.
\end{definition}

We would like to mention that for the above definition to be meaningful, the {\em displacement operators} $(\vL,\vR)$ need to satisfy some non-trivial requirements for $\vW$. E.g. if $\vL=\vR=\vI$, then all matrices have displacement rank $0$ with respect to $(\vI,\vI)$. However, if we insists that $\vL$ and $\vR$ do not share any common eigenvalues, then in the above definition, every $\vE$ corresponds to a unique matrix $\vW$. For the rest of the section, we will make this assumption.

\subsubsection{Some examples and arithmetic circuit complexity}

Consider the following matrix (called a Cauchy matrix):
\begin{definition}[Cauchy Matrix]
\label{def:chap2:cauchy}
Arbitrarily fix $\vs,\vt\in\F^n$ such that for every $0\le i,j<n$, $\vs[i]\neq\vt[j]$, $\vs[i]\neq \vs[j]$ and $\vt[i]\neq\vt[j]$ and
\[\vC_n[i,j]=\frac{1}{\vs[i]-\vt[j]}.\]
\end{definition}
It can be shown that this matrix has full rank. 
We next argue that the Cauchy matrix (Definition~\ref{def:chap2:cauchy}) has displacement rank $1$ with respect to $\vL=\diag(\vs)$ and $\vR=\diag(\vt)$, where recall $\diag(\vx)$ is the diagonal matrix with $\vx$ on its diagonal. Indeed, note that in this case we have $\diag(\vs)\vC_n-\vC_n\diag(\vt)$ is the all ones matrix. 

Further, it turns out that the Vandermonde matrix (Definition~\ref{def:chap1:vandermonde}) $\vV_n^{(\va)}$ for any $\va\in\F^n$ has displacement rank $1$ with respect to $\vL=\diag(\va)$ and $\vR$ being the {\em shift} matrix as defined next: 
\begin{definition}[Shift Matrix]
\label{def:chap5:shift}
The {\em shift} matrix $\vZ\in\F^{n\times n}$ is defined by
\[\vZ[i,j]=\begin{cases}
1&\text{ if } i=j-1\\
0&\text{ otherwise}.
\end{cases}
\]
\end{definition}
(The reason the matrix $\vZ$ is called the shift matrix is because when applied to the left or right of a matrix it shifts the row (or columns respectively) of the matrix.) 

It is known how to compute $\vW\vx$ with arithmetic circuit complexity $\tO{rn}$, where $\vW$ has displacement rank at most $r$ with respect to $\vL,\vR$ where these `operators' are either shift or diagonal matrices. 
In fact De Sa et al.~\cite{soda18} show that as long as $\vL$ and $\vR$ are $O(1)$-quasiseparable (i.e. all sub-matrices strictly above or strictly below the main diagonal are $O(1)$-rank) then any matrix $\vW$ that has rank $r$ with respect to $(\vL,\vR)$ has arithmetic circuit complexity of $\tO{rn}$.

\subsubsection{Low displacement rank matrices in deep learning literature}

Low displacement rank (or LDR) matrices have actually been implemented in deep learning systems with some success in reducing the memory footprint and the time efficiency of inference~\cite{sindhwani2015structured,ldr-neurips18}. 
Here we give a very quick (and necessarily incomplete) overview of the main results of paper of Zhao et al.~\cite{ldr-nn-pan}.

Zhao et al. consider LDR with respect to {\em any fixed} displacement operators $(\vL,\vR)$ as long as
\begin{itemize}
\item Both $\vL$ and $\vR$ are non-singular diagonalizable matrices,
\item $\vL^q=a\cdot \vI$ for some $1\le q\le n$ and non-zero $a\in\R$,
\item $\paren{\vI-a\vB^q}$ is non-singular, and
\item The eigenvalues of $\vR$ are distinct in absolute values.
\end{itemize}

Zhao et al. fix the displacement operators $(\vL,\vR)$ as above and consider one layer neural networks (as in our case) where the weight matrix $\vW$ has $O(1)$-displacement rank with respect to $(\vL,\vR)$.\footnote{This means that during the learning phase, we already know $\vL$ and $\vR$ and we only need to learn the residual.} Note that such matrices can be represented by just storing the residual $\vL\vW-\vW\vR$ and hence only needs $O(n)$ parameters overall. For the rest of this subsection, we will refer to these as LDR neural networks.

They show that for three well-studied properties of single layer neural networks, the one with LDR weight matrices as just as `good' as arbitrary weight matrices. Arguing these results formally is out of scope for this survey so here we just given a very high level informal statements (and refer the reader to the paper~\cite{ldr-nn-pan} for the formal statements and their proofs):
\begin{enumerate}
\item The {\em universal approximation theorem} states that an LDR neural network can approximate any continuous function to within arbitrary precision over any point (i.e. under $\ell_\infty$ error norm).
\item The paper also shows that for {\em any} probability distribution over an $n$-dimensional ball, an LDR neural network can approximate any function (w.r.t. the probability distribution) with squared error $O(1/n^2)$. A similar result was shown for neural networks with arbitrary weight matrix, which we have already seen needs $\Omega(n^2)$ parameters (while the LDR neural network only needs $O(n)$ parameters as observed above).
\item Zhao et al. also show that one can compute the required gradients for the gradient descent algorithm (where roughly speaking the complexity of computing the gradients depends on the arithmetic circuit complexity of $\vL$ and $\vR$).\footnote{At a high level this should not be surprising given the results in Section~\ref{sec:gradient}, though Zhao et al. do not utilize the generic connection we established in Section~\ref{sec:gradient}.}
\end{enumerate}

One practical drawback in this setup is that one fixes $\vL$ and $\vR$ upfront. Thomas et al. have run experiments where one tries to learn the displacement operator $\vL$ and $\vR$ {\em along} with the residual matrix~\cite{ldr-neurips18}.

\subsubsection{Coming back to Question~\ref{ques:main}}

Unfortunately, low displacement rank matrices are not enough to answer Question~\ref{ques:main} in the affirmative either. 

\begin{itemize}
\item (\propone) It turns out that the full power of low displacement rank matrices w.r.t. $O(1)$-quasiseparable displacement matrices is not known-- in other words, it is not known if these matrices satisfy \propone. We conjecture that they do {\em not}. In a somewhat weak support of this conjecture, we note that the traditional low displacement operators are either a (non-zero) diagonal matrix $\vD$ or (simple variants) of the shift matrix $\vZ$ (the initial experimental results on LDR neural networks are for these displacement operators~\cite{sindhwani2015structured})-- and in this case there are even diagonal matrices that have displacement rank $\Omega(n)$ with respect to these displacement operators (which means we have $s'\ge\Omega(n^2)$). 

Indeed, if at least one of $\vL$ or $\vR$ is a diagonal matrix, then we note that $\vL-\vR$ has all non-zero diagonal entries\footnote{If WLOG $\vL=\vZ$ and $\vR=\vD$, then the diagonal of $\vL-\vR$ is the diagonal of $\vD$ and hence all non-zero by our assumption. If both $\vL$ and $\vR$ are diagonal matrices, i.e. $\vL=\vD_1$ and $\vR=\vD_2$, then $\vL-\vR=\vD_1-\vD_2$ and all these entries are non-zero since we assumed $\vL$ and $\vR$ do not share any eigenvalues.} and is lower/upper triangular and hence $\vW=\vI$ has displacement rank of $\Omega(n)$ with respect to such matrices. If both $\vL$ and $\vR$ are shift matrices, then we note that for a diagonal matrix $\vD$, we have $\vE=\vZ\vD-\vD\vZ=\vD'\cdot\vZ$, where elements of $\vD'[i,i]=\vD[i,i]-\vD[i+1,i+1]$. Thus, if we choose $\vD$ such that all the consecutive elements on the diagonal are different, then we have that rank of $\vE$ is the same as rank of $\vZ$ and thus, $\vD$ will have displacement rank $\Omega(n)$ with respect to shift matrices.
\item (\proptwo) Results from~\cite{soda18} show that this property is satisfied when $\vL$ and $\vR$ are $O(1)$-quasiseparable matrices.
\item (\propthree) As mentioned earlier,~\cite{ldr-nn-pan} shows that this property is satisfied {\em if} $\vL$ and $\vR$ are fixed. Results in Section~\ref{sec:gradient} imply \propthree\ are satisfied as long as $\vW$ has efficient matrix-vector multiplication.
\end{itemize}

\subsection{Product of sparse matrices (in listing representation)}
\label{sec:spw}

All of the classes of structured matrices that we have considered so far have all not been able to satisfy \propone\ in Question~\ref{ques:main}. Next we consider the class of {\em product} of sparse matrices. De Sa et al.~\cite{soda18}, showed that these can accurately capture $\vW$ with small arithmetic circuits:
\begin{theorem}
\label{thm:spw}
     Let $\vW$ be an $n \times n$ matrix such that matrix-vector multiplication of $\vW$ times an arbitrary vector $\vv$ can be represented as a linear arithmetic circuit $C$ comprised of $s$ gates (including inputs) and having depth $d$. Then we can represent $\vW$ as a product of $d+1$ matrices each of which is $O(s)$ sparse.
\end{theorem}

In fact~\cite{soda18} also proves a `converse' of the above result (which means product of sparse matrices {\em exactly} capture the power of (linear) arithmetic circuits for linear maps).
Before we present the proof of the above result, we remark that Theorem~\ref{thm:spw} and its converse in~\cite{soda18}  are probably known but we have not been able to to find a reference that pre-dates~\cite{soda18}-- if you are aware of a reference for the above theorem, please let the author know.

\begin{proof}[Proof of Theorem~\ref{thm:spw}]
    We will represent $C$ as a product of $d$ matrices, each of size $s' \times s'$, where $s'$ is the smallest power of 2 that is greater than or equal to $s$.

    Define $w_1, \hdots w_d$ such that $w_k$ represents the number of gates in the $k$'th layer of $C$ (note that $s = n + \sum_{k=1}^{d}w_k$). Also, define $z_1, \hdots z_d$ such that $z_1 = n$ and $z_k = w_{k-1} + z_{k-1}$ ($z_k$ is the number of gates that have already been used by the time we get to layer $k$).

    Let $g_i$ denote the $i$'th gate (and its output) of $C$ ($0 \leq i < s$), defined such that (where we want to multiply $\vv=(v_0,\dots,v_{n-1})$ with $\vW$):
    \[
        g_i = \begin{cases}
            v_i & 0 \leq i < n \\
            \alpha_i g_{i_1} + \beta_i g_{i_2} & n \leq i < s
        \end{cases}
    \]

    where $i_1, i_2$ are indices of gates in earlier layers.

    For the $k$'th layer of $C$, we define the $s' \times s'$ matrix $\vW_k$ such that it performs the computations of the gates in that layer. Define the $i$'th row of $\vW_k$ to be:
    \[  
        \vW_k[i:] = 
        \begin{cases}
            \ve_i^T & 0 \leq i < z_k \\
            \alpha_{i} \ve_{i_1}^T + \beta_{i} \ve_{i_2}^T & z_k \leq i < z_k + w_k
            \\
            0 & i \geq z_k + w_k
        \end{cases}
    \]

    For any $0 \leq k \leq d$, let $\mathbf{\vv_k}$ be vector
    \[
        \vv_k = \vW_k \hdots \vW_2 \vW_1 \begin{bmatrix} \vv \\ \vzero \end{bmatrix}. 
    \]
    We'd like to argue that $\vv_d$ contains the outputs of all gates in $C$ (i.e, the $n$ values that make up $\vW\vv$). To do this we argue, by induction on $k$, that $\vv_k$ is the vector whose first $z_{k+1}$ entries are $g_0, g_1, \hdots, g_{(z_{k+1}-1)}$, and whose remaining entries are $0$. The base case, $k=0$ is trivial. Assuming this holds for the case $k-1$, and consider multiplying $\vv_{k-1}$ by $\vW_k$. The first $z_k$ rows of $\vW_k$ duplicate the first $z_k$ entries of $\vv_{k-1}$. The next $w_k$ rows perform the computation of gates $g_{z_k}, \hdots, g_{(z_{k+1}-1)}$. Finally, the remaining rows pad the output vector with zeros. Therefore, $\vv_k$ is exactly as desired.

    The final matrix product will contain all $n$ elements of the output, as desired. By left multiplying by some permutation matrix $\vP$, we can reorder this vector such that the first $n$ entries are exactly $\vW\vv$ (or more precisely we left multiply by a `truncated' permutation matrix so that the final answer is exactly $\vW\vv$). One can now check that we have product of $d+1$ matrices each of which is $O(s)$ sparse, as desired.
\end{proof}

We are now ready to evaluate whether product of sparse matrices can answer Question~\ref{ques:main} (spoiler alert: no!):

\begin{itemize}
\item (\propone) If we assume that we only consider $\vW$ that have circuit with depth $\tO{1}$ (which capture most of the known efficient matrix-vector multiplication algorithms), then Theorem~\ref{thm:spw} shows that $s'=O(ds)$, which by our assumption on $d$ is $\tO{s}$, which means we have satisfied \propone.
\item (\proptwo) If one uses the obvious algorithm (i.e. multiply successively by each of the $d+1$ matrices, each of which is $O(s)$-sparse), then one can compute the overall matrix vector multiplication in $O(ds)=O(s')$ operations. Thus, we also satisfy \proptwo.
\item (\propthree) This property is not satisfied if we assume the listing representation for each of the sparse matrices (due to the same reason that a single sparse matrix in listing representation does not satisfy \propthree).
\end{itemize}

We came close to answering Question~\ref{ques:main} with product of sparse matrices-- the only catch was that the listing representation of sparse matrices does not allow us to satisfy \propthree. Next, we answer Question~\ref{ques:main} in the positive by coming up with an alternative representation of sparse matrices that is differentiable.

\section{Butterfly matrices}
\label{sec:butterfly}
In this section, we will present a positive answer to Question~\ref{ques:main}.
We start with taking a circuit/matrix-product view of the FFT in Section~\ref{sec:fft}, which in turn motivates the definition of butterfly matrices in Section~\ref{sec:butterfly-def}. Finally, we use Butterfly matrices to define the final class of matrices in Section~\ref{sec:theory-main}, which we will show answer Question~\ref{ques:main} in the affirmative.

\subsection{Fast Fourier Transform (FFT)}
\label{sec:fft}

As mentioned earlier, a vast majority of efficient matrix vector multiplication algorithms are equivalent to small (both in size and depth) linear arithmetic circuit. For example the FFT can be thought of as an efficient arithmetic circuit to compute the Discrete Fourier Transform (indeed when one converts the linear arithmetic circuit for FFT into a matrix decomposition, 
then each matrix in the decomposition is   so called {\em Butterfly matrix}, with each block matrix in each factor being the same). For an illustration of this consider the DFT with $n=4$ as illustrated in Figure~\ref{fig:dft}.

\begin{figure}[H]
    \dftfigure
    \caption{DFT of order $4$.}
\label{fig:dft}
\end{figure}

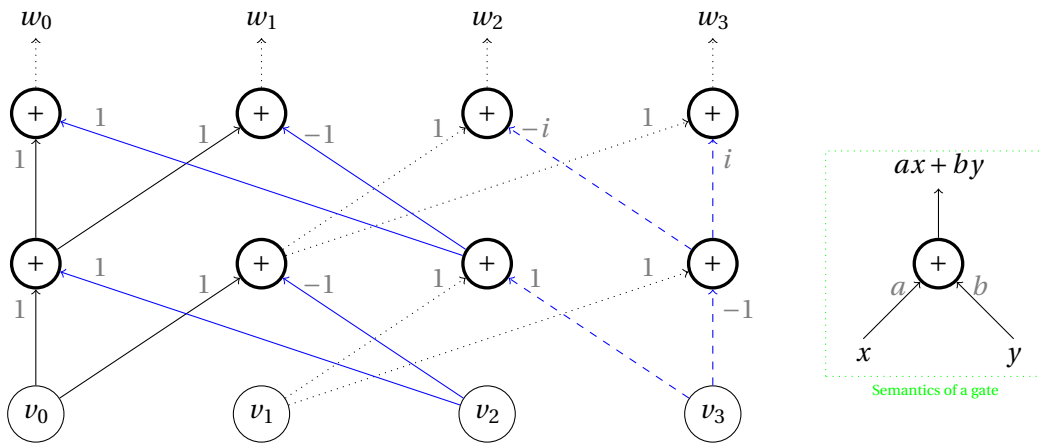
\begin{figure}[H]
    \begin{center}
    \begin{tikzpicture}

            \tikzset{vertex/.style = {shape=circle,draw,line width=1.25}}
            \tikzset{edge/.style = {->}}

    \node[shape=circle,draw] (v0) at  (0,0) {$v_0$};
    \node[shape=circle,draw] (v1) at  (3,0) {$v_1$};
    \node[shape=circle,draw] (v2) at  (6,0) {$v_2$};
    \node[shape=circle,draw] (v3) at  (9,0) {$v_3$};

    \node[vertex] (g0) at  (0,2) {$+$};
    \node[vertex] (g1) at  (3,2) {$+$};
    \node[vertex] (g2) at  (6,2) {$+$};
    \node[vertex] (g3) at  (9,2) {$+$};

    \node[vertex] (w0) at  (0,4) {$+$};
    \node[vertex] (w1) at  (3,4) {$+$};
    \node[vertex] (w2) at  (6,4) {$+$};
    \node[vertex] (w3) at  (9,4) {$+$};

    \draw[->] (v0) -- (g0) node[pos=0.8, left]{$\textcolor{gray}{1}$};
    \draw[->] (v0) -- (g1) node[pos=0.8, left, above]{$\textcolor{gray}{1}$};

    \draw[->, dotted] (v1) -- (g2) node[pos=0.85, left, above]{$\textcolor{gray}{1}$};
    \draw[->, dotted] (v1) -- (g3) node[pos=0.9, left, above]{$\textcolor{gray}{1}$};

    \draw[->, color=blue] (v2) -- (g0) node[pos=0.9, right, above]{$\textcolor{gray}{1}$};
    \draw[->, color=blue] (v2) -- (g1) node[pos=0.8, right, above]{$\textcolor{gray}{-1}$};

    \draw[->, color=blue, dashed] (v3) -- (g2) node[pos=0.85, right, above]{$\textcolor{gray}{1}$};
    \draw[->, color=blue, dashed] (v3) -- (g3) node[pos=0.8, right]{$\textcolor{gray}{-1}$};

    \draw[->] (g0) -- (w0) node[pos=0.8, left]{$\textcolor{gray}{1}$};
    \draw[->] (g0) -- (w1) node[pos=0.8, left, above]{$\textcolor{gray}{1}$};

    \draw[->, dotted] (g1) -- (w2) node[pos=0.85, left, above]{$\textcolor{gray}{1}$};
    \draw[->, dotted] (g1) -- (w3) node[pos=0.9, left, above]{$\textcolor{gray}{1}$};

    \draw[->, color=blue] (g2) -- (w0) node[pos=0.9, right, above]{$\textcolor{gray}{1}$};
    \draw[->, color=blue] (g2) -- (w1) node[pos=0.8, right, above]{$\textcolor{gray}{-1}$};

    \draw[->, color=blue, dashed] (g3) -- (w2) node[pos=0.85, right, above]{$\textcolor{gray}{-i}$};
    \draw[->, color=blue, dashed] (g3) -- (w3) node[pos=0.8, right]{$\textcolor{gray}{i}$};

    \draw[->,dotted] (w0) -- (0,5) node[pos=1, above] {$w_0$};
    \draw[->,dotted] (w1) -- (3,5) node[pos=1, above] {$w_1$};
    \draw[->,dotted] (w2) -- (6,5) node[pos=1, above] {$w_2$};
    \draw[->,dotted] (w3) -- (9,5) node[pos=1, above] {$w_3$};

    \node[vertex] (g) at  (12,2) {$+$};

    \draw[->] (11,1) -- (g) node[pos=0, below] {$x$} node[pos=.6, left , above]{$\textcolor{gray}{a}$};
    \draw[->] (13,1) -- (g) node[pos=0, below] {$y$} node[pos=.6, right , above]{$\textcolor{gray}{b}$};
    \draw[->] (g) -- (12,3) node[pos=1, above]{${ax+by}$};

    \draw[color=green, dotted] (10.5,0.5) rectangle (13.5,3.5);
    \node[right] (legend) at (11, 0.3) {\textcolor{green}{\tiny Semantics of a gate}};

\end{tikzpicture}
    \end{center}
    \caption{Arithmetic circuit for $4$-DFT from Figure~\ref{fig:dft}.}
\label{fig:dft-ac}
\end{figure}

Figure~\ref{fig:dft-ac} represent the arithmetic circuit corresponding to FFT with $n=4$.

\begin{figure}[H]
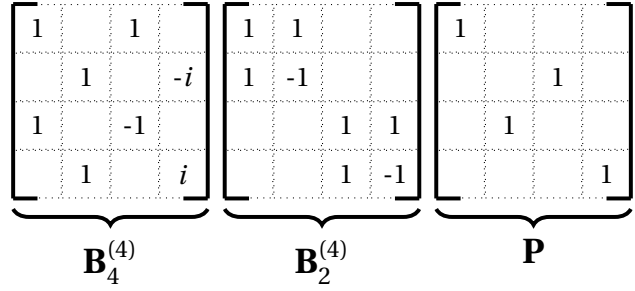

    \dftdecompfigure
    \caption{Decomposition of DFT of Figure~\ref{fig:dft} via the arithmetic circuit of Figure~\ref{fig:dft-ac}.}
\label{fig:dft-bbt}
\end{figure}

Finally, Figure~\ref{fig:dft-bbt} is representation of the arithmetic circuit of Figure~\ref{fig:dft-ac} as a product of a butterfly matrix and (the bit-reversal) permutation. 

\subsection{Butterfly matrices}
\label{sec:butterfly-def}

Butterfly matrices, encoding the recursive divide-and-conquer structure of
the fast Fourier transform (FFT) algorithm as illustrated in Figure~\ref{fig:dft-bbt}, have long been used in numerical linear algebra~\cite{parker1995random, li2015butterfly} and machine learning~\cite{mathieu2014fast, jing2017tunable, munkhoeva2018quadrature, dao2019learning, choromanski2019unifying}.
Here we define butterfly matrices, which we use as a building block for our hierarchy of kaleidoscope matrices.

\begin{definition} \label{def:bfactor}
    A \textbf{butterfly factor} of size $k\ge 2$ (denoted as $\vB_k$) is a matrix of the form
    \(
        \vB_k = \begin{bmatrix}
            \vD_1 & \vD_2 \\ \vD_3 & \vD_4
        \end{bmatrix}
    \)
    where each $\vD_i$ is a $\frac{k}{2} \times \frac{k}{2}$ diagonal matrix. We restrict $k$ to be a power of 2.
\end{definition}

\begin{definition}
     A \textbf{butterfly factor matrix} of size $n$ with block size $k$ (denoted as $\vB_k^{(n)}$) is a block diagonal matrix of $\frac{n}{k}$ (possibly different) butterfly factors of size $k$:
     \[
        \vB_k^{(n)} = \mathrm{diag} \left( \left[ \vB_k \right]_1, \left[ \vB_k \right]_2, \hdots, \left[ \vB_k \right]_\frac{n}{k} \right)
     \]
\end{definition}

\begin{definition} \label{def:bmatrix}
    A \textbf{butterfly matrix} of size $n$ (denoted as $\vB^{(n)}$) is a matrix that can be expressed as a product of butterfly factor matrices:
    \(
        \vB^{(n)} = \vB_n^{(n)} \vB_{\frac{n}{2}}^{(n)} \hdots \vB_2^{(n)}.
    \)
    Equivalently, we may define $\vB^{(n)}$ recursively as a matrix that can be expressed in the following form:
    \[
         \vB^{(n)} = \vB_n^{(n)} \begin{bmatrix}
            [\vB^{(\frac{n}{2})}]_1 & 0 \\
            0 & [\vB^{(\frac{n}{2})}]_2
         \end{bmatrix}
    \]
    (Note that $[\vB^{(\frac{n}{2})}]_1$ and $[\vB^{(\frac{n}{2})}]_2$ may be different.)
\end{definition}

\subsection{The kaleidoscope hierarchy}
\label{sec:theory-main}


Using the building block of butterfly matrices, we formally define the kaleidoscope
($\BBS$) hierarchy and prove its expressiveness.
This class of matrices serves as a fully differentiable alternative to products of sparse matrices
(Section~\ref{sec:spw}), with similar expressivity. This family of matrices was defined by Dao et al.~\cite{k-mat}.

\begin{figure}[th]
  \centering
  \includegraphics[width=1.0\linewidth]{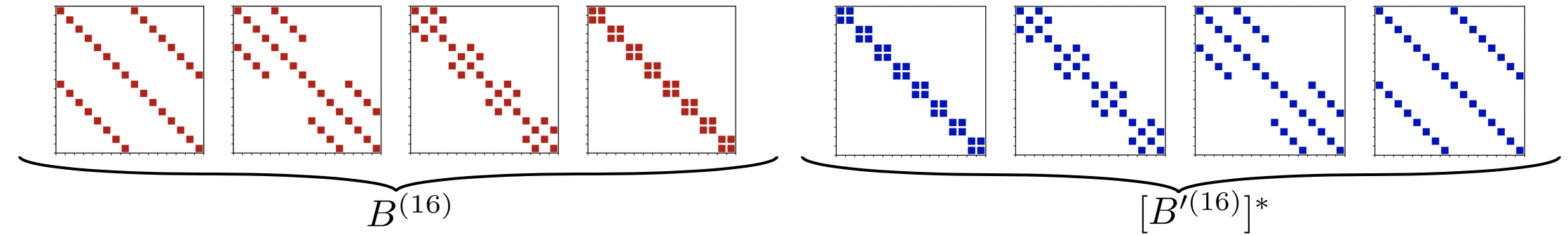}
  \caption{\label{fig:bbs_sparsity} Visualization of the fixed sparsity pattern
    of the building blocks in $\BBS$, in the case $n = 16$. The red and blue
    dots represent all the possible locations of the nonzero entries.}
\end{figure}

The building block for this hierarchy is the product of a butterfly matrix and
the (conjugate) transpose of another butterfly matrix (which is simply a product
of butterfly factors taken in the opposite order).
Figure~\ref{fig:bbs_sparsity} visualizes the sparsity patterns of the
butterfly factors in $\BBS$, where the red and blue dots represent the
allowed locations of nonzero entries.

We would like to note that the sparsity pattern in a matrix in $\BBS$ matches {\em exactly} the Bene\v{s} network~\cite{benes-jour,benes-book}, which is a multistage circuit switching network. The goal in a switching network in Bene\v{s} network is to route $n$ input connection to $n$ output connection through a sequence of switches where the basic building block is a {\em cross-bar switch} (where each such switch can `swap' two connections). It is known that the Bene\v{s} network can route {\em any permutation} from the input connection to the output connection by appropriately making the switch swap (or not) its two input connections. In our setup of $\BBS$, we allow each `switch' in a Bene\v{s} network to be replaced by an arbitrary $2\times 2$ sub-matrix.

\begin{definition}[Kaleidoscope hierarchy, kaleidoscope matrices]\ \label{def:hierarchy}
    \begin{itemize}
        \item Define $\B$ as the set of all matrices that can be expressed in the form $\vB^{(n)}$ (for some $n$).
        \item Define $\B\B^*$ as the set of matrices $\vM$ of the form $\vM = \vM_1 \vM_2^*$\, for some $\vM_1, \vM_2 \in \B$.
        \item Define $(\BBS)^w$ as the set of matrices $\vM$ that can be expressed as $\vM = \vM_w \hdots \vM_2 \vM_1$, with each $\vM_i \in \BBS$ ($1 \leq i \leq w$). (The notation $w$ represents $\textbf{width}$.)
        \item Define $(\BBS)^w_e$ as the set of $n \times n$ matrices $\vM$ that can be expressed as $\vM  = \vS \vE \vS^T$ for some $en \times en$ matrix $\vE \in (\BBS)^w$, where $\vS \in \mathbb{F}^{n \times en} = \begin{bmatrix} \vI_n & 0 & \hdots & 0 \end{bmatrix}$ (i.e. $\vM$ is the upper-left corner of $\vE$). (The notation $e$ represents \textbf{expansion} relative to $n$.)
        \item $\vM$ is a \textbf{kaleidoscope matrix}, abbreviated as
        \textbf{K-matrix}, if $M \in (\BBS)^w_e$ for some $w$ and $e$.
    \end{itemize}
\end{definition}\label{sec:hierarchy-def}
The \emph{kaleidoscope hierarchy}, or $(\BBS)$ hierarchy, refers to the families of
matrices $(\BBS)^1_e \subseteq (\BBS)^2_e \subseteq \dots$, for a fixed expansion factor $e$.
Each butterfly matrix can represent the identity matrix, so
$(\BBS)^w_e \subseteq (\BBS)^{w+1}_e$.
Dao et al.~\cite{k-mat} show that the inclusion is proper. 


\paragraph{Efficiency in space and speed.} Each matrix in $(\BBS)^w_e$ is a
product of $2w$ total butterfly matrices and transposes of butterfly matrices,
each of which is in turn a product of $\log (ne)$ factors with $2ne$ nonzeros (NNZ) each.
Therefore, each matrix in $(\BBS)^w_e$ has $4w ne \log (ne)$ parameters and a
matrix-vector multiplication algorithm of complexity $O(w ne \log ne)$ (by
multiplying the vector with each sparse factor sequentially).

\paragraph{Difference from family of matrices in Section~\ref{sec:spw}.} We note that the K-matrices are similar to the family of matrices considered in Section~\ref{sec:spw} in that they are also product of sparse matrices. The main difference is that each matrix in the product in addition to being sparse is also {\em structured}-- i.e. we know upfront where all the non-zero elements in each factor in a K-matrix will be. This allows us to create a differentiable representation for sparse matrices, which was the missing part of the family of product of (general) sparse matrices.

\subsubsection{Answering Question~\ref{ques:main}}

We state the main theoretical result, namely, the ability to capture general transformations, expressed as low-depth linear arithmetic circuits, in the $\BBS$ hierarchy. This result is recorded in Theorem~\ref{thm:ac-bbs}.


 \begin{theorem} \label{thm:ac-bbs}
     Let $\vM$ be an $n \times n$ matrix such that matrix-vector multiplication of $\vM$ times an arbitrary vector $\vv$ can be represented as a linear arithmetic circuit $C$ comprised of $s$ gates (including inputs) and having depth $d$. Then, $\vM \in (\BBS)^{O(d)}_{O\paren{\frac{s}{n}}}$.
\end{theorem}

Before we prove Theorem~\ref{thm:ac-bbs}, we note that it is sufficient to show that K-matrices answer Question~\ref{ques:main} in the affirmative:
\begin{enumerate}
\item (\propone) Theorem~\ref{thm:ac-bbs} along with the observation on number of parameters needed to represent a matrix in $(\BBS)^w_e$ implies that we have $s'=O\paren{d\cdot \frac sn\cdot n\log\paren{\frac sn\cdot n}}=O(ds\log{s})$. Thus, under the assumption of $d=\tO{1}$, we have that $s'=\tO{s}$, as desired.
\item (\proptwo) Again by Theorem~\ref{thm:ac-bbs} along observation on number of operations needed to do matrix-vector multiplication for a matrix in $(\BBS)^w_e$ (and using the calculations from the previous bullet), we get that the matrix-vector multiplication takes $O(s')$ operations, as desired.
\item (\propthree) Finally, since we know the locations of the non-zero elements (which form the parameters for K-matrices), it is not too hard to see that each entry in $\vW_\vtheta$ is a polynomial in the entries of $\vtheta$. Since a polynomial in $\vtheta$ is differentiable, this means \propthree\ is satisfied as well.
\end{enumerate}

\paragraph{Proof of Theorem~\ref{thm:ac-bbs}.}

To prove Theorem~\ref{thm:ac-bbs}, we make use of the following two theorems.

\begin{theorem} \label{thm:perm}
    Let $\vP$ be an $n \times n$ permutation matrix (with $n$ a power of 2). Then $\vP \in \BBS$.
\end{theorem}

\begin{theorem} \label{thm:sparse}
    Let $\vS$ be an $n \times n$ matrix of $s$ NNZ. Then $\vS \in (\BBS)_4^{4\ceil{\frac{s}{n}}}$.
\end{theorem}

We first give an overview of how the above two results imply Theorem~\ref{thm:ac-bbs} and then briefly outline how one can prove the two results above. First, we note that (proof of) Theorem~\ref{thm:spw} implies that given any $\vW$ with arithmetic circuit of size $s$ and depth $d$, we can represent $\vW$ as a product of $d$ many $O(s)$-sparse matrices and a permutation matrix. Thus, we can decompose $\vW$ as product of $O(d)$ K-matrices. Then Theorem~\ref{thm:ac-bbs} follows from the simple observation that membership in the family of K-matrices is closed under multiplication.

Theorem~\ref{thm:perm} essentially follows from the known fact that a Bene\v{s} network can route an arbitrary permutation (see~\cite{k-mat} for a self-contained proof in the language of K-matrices).

Theorem~\ref{thm:sparse} follows by showing that any $n$-sparse matrix is in $(\BBS)^4$ and the fact that membership in the family of K-matrices is closed under addition. To show the inclusion of an $n$-sparse matrix, Dao et al.~\cite{k-mat} show that any $n$-sparse matrix $\vS$ can be decomposed as $\vP_1\vH\vP_2\vV\vP_3$, where $\vP_1,\vP_2,\vP_3$ are permutation matrices (which by Theorem~\ref{thm:perm} are in $\BBS$). $\vH$ is {\em horizontal step matrix}, which obeys a `Lipschitz-like' condition. Each column of a horizontal step matrix can have at most one non-zero entry, and given two non-zero columns $k$ apart, the non-zero entry in the right column must be between 0 and $k$ rows below the non-zero entry in the left column. Note that to show that a matrix is a horizontal step matrix, it is sufficient to argue that this condition holds for each pair of neighboring non-zero columns. The matrix $\vV$ is such that its transpose is a horizontal step matrix. Dao et al.~\cite{k-mat} show that any horizontal step matrix is in $\B$. Combining all of these, we have that $\vS\in (\BBS)(\B)(\BBS)(\B^*)(\BBS)\subseteq (\BBS)^5$. Dao et al.~\cite{k-mat} observe that with bit more careful analysis we can show inclusion in $(\BBS)^4$. We refer the interested reader to~\cite{k-mat} for the proof details.

\section{Open Questions}
\label{sec:concl}

We conclude by present two open questions (the first one being a specific technical question and the other one being a bit more vague):
\begin{enumerate}
\item There is one unsatisfactory aspect to the results in Section~\ref{sec:theory-main}, i.e. the number of parameters needed to specify the family of K-matrices that capture matrices with arithmetic circuit of size $s$ and depth $d$ is $O(sd\log{s})$. In particular, the dependence on $d$ is not ideal, which leads to the following:
\begin{oques}
Is it possible to answer Question~\ref{ques:main} in the affirmative with a family that uses $s'=\tO{s}$ many parameters to capture all matrices with arithmetic circuits of size $s$ (irrespective of the depth $d$)?
\end{oques}
\item As mentioned earlier, low rank approximation is ubiquitous in machine learning (and numerical linear algebra more generally). One intriguing possibility is whether K-matrices can replace low rank matrices in these applications? Currently, the main technical stumbling block is solving the following:
\begin{oques}
Does there exist an efficient algorithm that solves the following problem-- given an arbitrary matrix $\vM\in\F^{n\times n}$ and parameters $w$ and $e$, find the matrix $\vW\in (\BBS)^w_e$ that is closest (or `close enough') to $\vM$ (say in Frobenius norm)?
\end{oques}
We note that for low rank matrices, the SVD solves the above question. Thus, the question is asking whether we can be design the `SVD for K-matrices'? Partial progress on a variant of the above question was made recently in~\cite{monarch}.
\end{enumerate}

\subsection*{Acknowledgments}

The material in Sections~\ref{sec:prelim} and~\ref{sec:gradient} are based on notes for AR's Open lectures for PhD students in computer science at University of Warsaw titled {\em (Dense Structured) Matrix Vector Multiplication} in May 2018-- we would like to thank University of Warsaw's hospitality. The material in Section~\ref{sec:butterfly} is based on Dao et al.~\cite{k-mat}.

We would like to thank Tri Dao, Albert Gu and Chris R\'{e} for many illuminating discussions during our collaborations around these topics.

We would like to thank an anonymous reviewer whose comments improved the presentation of the survey (and for pointing us to Theorem~\ref{thm:vand-lb}) and we thank Jessica Grogan for a careful read of an earlier draft of this survey.

AR is supported in part by NSF grant CCF-1763481.

\bibliographystyle{plain}
\bibliography{survey,butterfly_iclr_2020}


\end{document}